\documentclass[twocolumn,aps,prd,superscriptaddress,preprintnumbers,nofootinbib,10pt]{revtex4-2}
\usepackage{amsmath,amssymb}
\usepackage[dvipdf,dvips]{graphicx}
\usepackage{color}
\usepackage{hyperref}
\usepackage{url}
\usepackage{slashed}
\usepackage{subfigure}
\usepackage[usenames,dvipsnames]{xcolor}
\usepackage{amsmath}
\usepackage{amsfonts}
\usepackage{float} 
\usepackage{amssymb}
\usepackage{epsfig}
\usepackage{graphics}
\usepackage{euscript}
\usepackage{slashed}
\usepackage{epstopdf}
\usepackage[utf8]{inputenc}
\allowdisplaybreaks

\hypersetup{
colorlinks=true,
citecolor=blue,
citebordercolor=red,
linktoc=all,
linkcolor=blue,
urlcolor=blue
}

\def \and{\textmd{and}}

\def \be{\begin{equation}}
\def \ee{\end{equation}}

\def \bea{\begin{eqnarray}}
\def \eea{\end{eqnarray}}

\newbox{\ORCIDicon}
\sbox{\ORCIDicon}{\large
                  \includegraphics[width=0.8em]{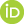}}

\begin{document}

\title{Beyond gauge-parameter dependence: Further evidence for \\
asymptotic safety in Scalar-Tensor theories}

\author{Samuel M. P. Antunes} \email{samuelmotta@id.uff.br}
\affiliation{Instituto de F\'isica, Universidade Federal Fluminense, Campus da Praia Vermelha, Av. Litor\^anea s/n, 24210-346, Niter\'oi, RJ, Brazil}

\author{Antonio D. Pereira\,\href{https://orcid.org/0000-0002-6952-2961}{\usebox{\ORCIDicon}}} \email{adpjunior@id.uff.br}
\affiliation{Instituto de F\'isica, Universidade Federal Fluminense, Campus da Praia Vermelha, Av. Litor\^anea s/n, 24210-346, Niter\'oi, RJ, Brazil}

\begin{abstract}
Asymptotic Safety provides a promising framework for the quantization of Scalar-Tensor theories as non-perturbatively renormalizable quantum field theories. Over the past years, growing evidence has accumulated for the existence of non-trivial renormalization group fixed points in a variety of Scalar-Tensor models. In this work, we investigate the robustness of these fixed points under significant modifications of the computational setup. In particular, we employ a gauge condition that explicitly depends on the dynamical scalar field, going beyond a mere gauge-parameter deformation of the standard linear covariant gauges commonly used in the literature. Exploring different field parameterizations, we identify fixed points belonging to the Gaussian Matter class in four spacetime dimensions that are consistent with previous studies. The real part of the critical exponents associated with the scalar-field mass and the non-minimal scalar-curvature coupling turn out to have very small absolute value. We provide a comprehensive exploration of how the anomalous dimension of the scalar field and field parameterization impact those critical exponents showing that fixed-point candidates with two or four relevant directions emerge. 
\end{abstract}

\maketitle

\section{Introduction \label{Sec:Intro}}

The perturbative non-renormalizability of General Relativity \cite{tHooft:1974toh,Christensen:1979iy,Goroff:1985th} is often interpreted as a manifestation of the incompatibility between Quantum Field Theory (QFT) and gravity. Searching for different strategies to construct a quantum theory for the gravitational interaction has led to a significant proliferation of quantum-gravity (QG) approaches, see, e.g., \cite{deBoer:2022zka,Buoninfante:2024yth} and \cite{Basile:2024oms}. However, regarding QFT and gravity as incompatible can be problematic, at least, in two aspects: First, it is possible to develop a perfectly well-defined QFT for the gravitational field in the context of effective field theory. Such a program leads to concrete predictions for quantum-gravity corrections, see, e.g., \cite{Donoghue:1993eb,Donoghue:1994dn,Bjerrum-Bohr:2014zsa,Bjerrum-Bohr:2016hpa}. However, such an effective description breaks down near the Planck scale. Secondly, perturbative renormalizability is neither necessary nor sufficient to define a consistent quantum theory valid across arbitrary energy scales. A paradigmatic example is Quantum Electrodynamics (QED). It is indisputable that QED is a very successful theory with impressively accurate predictions. It is perturbatively renormalizable. Yet, the theory features a Landau pole, i.e., its coupling diverges at finite energy scales, see \cite{Gell-Mann:1954yli,Gockeler:1997dn}. The same happens with standard four-dimensional scalar field theory, see \cite{Frohlich:1982tw}. This leads to the conclusion that perturbative renormalizability does not guarantee the validity of a QFT at arbitrarily high energies. One possibility to ensure a well-defined behavior of a QFT as energy grows is the existence of a renormalization-group (RG) fixed point (FP). The running of the gauge coupling in Quantum Chromodynamics is such that it becomes asymptotically free in the ultraviolet (UV). It means that the coupling runs to zero, i.e., it hits a FP at vanishing coupling and the theory becomes free in the deep UV. In this case, the FP is said to be trivial or Gaussian. Due to the FP nature, perturbation theory is sufficient to probe its existence. Alternatively, the QFT might have a UV limit where the underlying FP is not trivial and the theory remains interacting even after the couplings reach the FP. In this case, the theory is said to be asymptotically safe and the FP is named non-trivial or non-Gaussian. Since the FP can be sufficiently far from the perturbative regime (i.e., sufficiently large), it is, in general, necessary to apply non-perturbative techniques to investigate its existence. In \cite{Hawking:1979ig}, Weinberg conjectured that QG could be described by an asymptotically safe QFT. This approach to QG has developed significantly, leading to the asymptotic safety scenario for QG or, simply, asymptotically safe QG (ASQG). The systematic investigation of the non-Gaussian FP (NGFP) structure in ASQG was made possible after the groundbreaking work by Reuter \cite{Reuter:1996cp}. Technically, an exact renormalization group equation was developed for gravity, allowing for non-perturbative computation of the corresponding beta functions. This was implemented in the framework of the Functional Renormalization Group (FRG) \cite{Wetterich:1992yh,Morris:1993qb,Berges:2000ew,Dupuis:2020fhh} adapted for gauge theories \cite{Reuter:1993kw,Ellwanger:1994iz,Ellwanger:1995qf,Pawlowski:2005xe,Gies:2006wv}. See \cite{Reuter:2007rv,Reuter:2012xf,Percacci:2017fkn,Reuter:2019byg,Reichert:2020mja} for pedagogical introductions to ASQG.

Despite being an exact equation, extracting concrete results from the FRG equation (or, simply, flow equation) requires approximations (truncations). They do not need to rely on perturbative assumptions, and that is the reason why it is possible to obtain non-perturbative information. Their quality is assessed by the (apparent) convergence of results upon refinement of the underlying truncation. In ASQG, compelling evidence for the existence of an NGFP, also called Reuter FP, has been found over the past years \cite{Souma:1999at,Reuter:2001ag,Litim:2003vp,Codello:2006in,Benedetti:2009rx,Benedetti:2009gn,Manrique:2010am,Manrique:2011jc,Christiansen:2012rx,Falls:2013bv,Benedetti:2013jk,Codello:2013fpa,Falls:2014tra,Christiansen:2014raa,Christiansen:2015rva,Gies:2015tca,Gies:2016con,Biemans:2016rvp,Christiansen:2016sjn,Denz:2016qks,Knorr:2017fus,Knorr:2017mhu,Christiansen:2017bsy,Falls:2017lst} thanks to sophisticated improvements in truncations. See, e.g., \cite{Niedermaier:2006wt,Codello:2008vh,Percacci:2011fr,Reuter:2012id,Eichhorn:2017egq,Eichhorn:2018yfc,Wetterich:2019qzx,Pereira:2019dbn,Eichhorn:2020mte,Bonanno:2020bil,Pawlowski:2020qer,Eichhorn:2022jqj,Eichhorn:2022gku,Saueressig:2023irs,Pawlowski:2023gym,Eichhorn:2026uqj} for some reviews on ASQG. One can use different non-perturbative techniques to search for a suitable NGFP in QG such as lattice simulations, see, e.g., \cite{Loll:2019rdj,Ambjorn:2020rcn,Ambjorn:2024pyv,Ambjorn:2024qoe,Laiho:2016nlp} and semi-analytical tools to describe the continuum limit of discrete models, see, e.g., \cite{Eichhorn:2017xhy,Eichhorn:2018ylk,Eichhorn:2018phj,Eichhorn:2019hsa,Castro:2026zch}. The inclusion of matter is not only relevant for phenomenological purposes but also from a structural viewpoint. Quantum fluctuations arising from matter fields might affect the FP structure for the gravitational couplings and, eventually, destroy their asymptotically safe behavior. Conversely, quantum-gravity fluctuations can affect the RG flow of matter couplings. After all, since one wants a consistent QFT describing the quantum nature of spacetime that is compatible with the matter content we know to exist, understanding the behavior of the full gravity-matter system is essential. 

Most of the development in ASQG is based on a QFT for the metric. Moreover, it is, in general, imposed that the underlying QFT is diffeomorphism invariant\footnote{This must be understood within the context of gauge theories, i.e., a gauge-fixing term might be included replacing gauge invariance by BRST invariance.}, thus constraining the operators that parameterize the effective action. The set of essential operators, i.e., those that cannot be eliminated by field redefinitions, forms a basis for the so-called theory space. By changing the underlying symmetry and/or the field content, one explores different theory spaces. Hence, by exploring the UV behavior of each QFT, one can probe different candidates for a QFT for the gravitational field if different UV completions exist. It is also logically possible that seemingly different theories actually fall in the same universality class. In particular, this strategy allows for searches of UV completions of modified theories of gravity. In this work, we explore Scalar-Tensor Theories (STT) which, besides the metric, contains a scalar field as a fundamental degree of freedom. The first investigations on the possible role of scalar fields in gravitational theories
date back to the works by Jordan, Brans, and Dicke \cite{jordan1948five,jordan1959present,Brans:1961sx,brans1961mach}. The central idea behind STTs is the introduction of a fundamental
scalar field that becomes one of the mediators of the gravitational interaction, alongside the graviton. Compelling motivations for studying STTs are the role of scalar fields for early- and late-time expansion of the Universe and their potential connection to dark matter \cite{Guth:1980zm,Linde:1981mu,Wetterich:1987fm,Bassett:2005xm,Copeland:2006wr,Tsujikawa:2013fta,Silveira:1985rk,McDonald:1993ex,Burgess:2000yq,Boehm:2003hm,Ferreira:2020fam}.

 In the context of asymptotic safety, STTs (or, equivalently, gravity-scalar systems) are widely explored. Besides the potential implications of an asymptotically-safe fixed point to inflation and dark-matter physics, the repercussions of quantum-gravity fluctuations to the Higgs sector of the Standard Model (and vice-versa) are at the heart of key developments in ASQG. See, e.g., \cite{Percacci:2003jz,Narain:2009fy,Narain:2009gb,Shaposhnikov:2009pv,Zanusso:2009bs,Vacca:2010mj,Eichhorn:2012va,Henz:2013oxa,Percacci:2015wwa,Labus:2015ska,Wetterich:2016uxm,Oda:2015sma,Henz:2016aoh,Eichhorn:2016esv,Hamada:2017rvn,Eichhorn:2017als,Rubio:2017gty,Pawlowski:2018ixd,Wetterich:2019zdo,Kwapisz:2019wrl,Reichert:2019car,Wetterich:2019rsn,Eichhorn:2020kca,Eichhorn:2020sbo,Ali:2020znq,deBrito:2021pyi,Laporte:2021kyp,Ohta:2021bkc,deBrito:2021akp,Eichhorn:2021tsx,Knorr:2022ilz,deBrito:2023myf,Wetterich:2024ieb,Ohta:2025xxo,Assant:2025gto,deBrito:2025nog,Bonanno:2025qsc,Maitiniyazi:2025pou,Bonanno:2026mzs,Bonanno:2026aiw,Bonanno:2026ljx,Wetterich:2026lce}.

In this work, we provide additional evidence for the existence of a RG-fixed point in STTs. Our strategy is to explore gauge- and field-parameterization dependence of FP solutions of the FRG within STTs. In particular, we focus on a gauge condition that involves the (dynamical) scalar field itself in contrast to most of the literature (see, however, \cite{Ohta:2025xxo}). The main reason for this choice is the fact that it renders a gauge-fixing action that eliminates non-minimal terms from the Hessian, allowing for a straightforward use of heat-kernel techniques. This strategy is widely employed in perturbative computations of quantum corrections in QG by means of the background-field method. However, most of the developments in the asymptotic-safety literature with a view towards STTs are based on the standard two-parameters family of linear covariant de Donder-like gauges together with a York decomposition, see, e.g., \cite{Narain:2009fy}. Such a procedure allows for the investigation of gauge-parameter dependence. Nevertheless, the universal properties of a genuine FP should hold beyond that. In fact, choosing a different gauge condition that is not just a gauge-parameter deformation of a previously chosen gauge allows for a complementary test of gauge dependence of the FP properties. Due to the limitations of truncated computations, there is little reason to expect that the underlying FP analysis will reveal similar properties with respect to the other class of gauges unless the FP is genuine and not a truncation/gauge artifact. Moreover, other arbitrary choices than gauge condition such as field parameterization can play a significant role in such an investigation. In practice, some properties of a would-be FP solution can change not only quantitatively but also qualitatively under different choices of gauge. However, if the FP is genuine, then it is logically conceivable that (at least part of) such modifications can be compensated by different choices of, e.g., field parameterization. Ultimately, a comprehensive investigation of all such spurious dependencies is crucial to establish the robustness of FP solutions obtained from truncated computations within the FRG. We have found that, within the approximations employed in this work, a suitable FP candidate that is compatible with the existing literature is found. The FP candidates belong to the class of Gaussian Matter FP (GMFP) in which the gravitational couplings attain non-trivial values while all the couplings associated with the scalar field vanish, see \cite{Narain:2009fy}. Notably, the FP candidates that we have found feature either two or four relevant directions. The reason for such an apparent discrepancy is the following: The real parts of the critical exponents associated with the scalar-field mass and the scalar-curvature coupling have very small absolute value. Since it is the sign of the real part of a given critical exponent that decides the relevance of the underlying coupling, the correct counting of relevant directions depends on the quantitative precision that will tell whether those real parts are slightly above or below zero. In this work, we are not able to establish that. However, within the approximations that we have employed, we have verified that those couplings are nearly marginal at the GMFP. 

This work is organized as follows: In Sect.~\ref{Sec:FRG}, we provide a short overview of the FRG and its application to QG mostly to fix our notation and conventions; In Sect.~\ref{Sec:STT}, we introduce the class of STTs we explore in this work together with the (minimal) gauge condition we employ and the corresponding Faddeev-Popov ghost sector. Sect.~\ref{Sect:Hessian} is devoted to the discussion of the regularization of the Hessian and the definition of the underlying projection rules that are employed for the extraction of the beta functions. In Sect.~\ref{Sect:FP4d}, we analyze the FP solutions obtained in four dimensions in the absence of the scalar-field anomalous dimension which is included in Sect.~\ref{Sect:FP4dAD}. In Sect.~\ref{Sect:ParametricAnalysis}, we analyze the critical exponents associated with the scalar-field mass and the non-minimal scalar-curvature which play a central role in the determination of the dimensionality of the critical surface. Finally, we present our conclusions and perspectives.

\section{Functional Renormalization Group: Key Elements\label{Sec:FRG}}

In this section, we provide a short review of the FRG, emphasizing the main aspects that are relevant for the purposes of this work. We divide the presentation in two parts. In the first one, we discuss the general features of the FRG while in the second we focus on the particular aspects of the application of the FRG to gravity. For more details, we refer to \cite{Morris:1998da,Berges:2000ew,Aoki:2000wm,Bagnuls:2000ae,Polonyi:2001se,Pawlowski:2005xe,Gies:2006wv,Delamotte:2007pf,Rosten:2010vm,Braun:2011pp,Dupuis:2020fhh,Eichhorn:2026uqj}.

\subsection{General Ideas}

The Wilsonian philosophy of the RG is characterized by an integration of quantum fluctuations in a (momentum) shell-wise way from a path-integral perspective. The FRG implements such procedure by means of a modification to the Boltzmann weight of the Euclidean path integral. A momentum-dependent mass-like term, the regulator, is introduced in order to suppress field modes with momentum lower than a given (unphysical) scale $k$. Such a scale works as an infrared (IR) cutoff. As $k$ is lowered, more and more fluctuations are integrated out. At $k=0$, all modes are taken into account, i.e., the functional integral is fully performed. At each value of $k$, one can define the so-called effective average action (EAA) $\Gamma_k$ which coincides with the effective action (i.e., the generating functional of one-particle irreducible diagrams), namely $\Gamma_{k=0} \equiv \Gamma$. If $k$ is chosen at a UV cutoff $\Lambda$ which might or might not be taken to infinity, no quantum fluctuations are integrated out and $\Gamma_{k=\Lambda}$ coincides with the classical action $S$, i.e., the one that is introduced in the Boltzmann weight. The EAA satisfies an exact equation, the Wetterich equation, which reads
\begin{equation}
\partial_t \Gamma_k = \frac{1}{2}\mathrm{STr}\left[\left(\Gamma^{(2)}_k + \mathcal{R}_k\right)^{-1}\partial_t \mathcal{R}_k\right]\,.
\label{Eq:FRG.1}
\end{equation}
The scale-derivative is defined as $\partial_t := k\partial_k$, $\mathrm{STr}$ represents a supertrace which, besides the sum over internal and spacetime indices, the integral over momenta and spacetime, also introduces a factor of $2$ and $-2$ if the corresponding fields are complex bosonic or Grassmmanian, respectively. The symbol $\Gamma^{(2)}_k$ is a short-hand notation for the Hessian, i.e., the matrix containing second-derivatives of the EAA with respect to all possible combinations of the fields present in the theory. Finally, $\mathcal{R}_k$ is the matrix containing all the regulator functions introduced through the regulator added in the Boltzmann weight in order to suppress modes with momentum below $k$. In order to have the flow equation with the simple one-loop structure as in \eqref{Eq:FRG.1} the regulator must be quadratic in the fields. Therefore, the regulator has the following form,
\begin{equation}
\Delta S_k = \frac{1}{2}\int {\mathrm{d}^dx}~\Phi^A (x) \mathcal{R}^{AB}_k (-\partial^2) \Phi^B (x)\,,
\label{Eq:FRG.2}
\end{equation}
which is a momentum-dependent mass-like term. The field content is represented collectively as $\Phi^A$ and the capital latin indices collect internal and spacetime indices. Solving Eq.\eqref{Eq:FRG.1} and flowing down to $k=0$ is equivalent to fully perform the path integral. However, as previously mentioned, Eq.\eqref{Eq:FRG.1} cannot be solved exactly and approximations are required. The EAA can be parameterized as 
\begin{equation}
\Gamma_k [\Phi] = \sum_i \bar{g}_i \EuScript{O}_i [\Phi]\,,
\label{Eq:FRG.3}
\end{equation}
with $\EuScript{O}_i [\Phi]$ being an operator built out of the fields $\Phi^A$ which is compatible with the symmetries of the classical action $S$ deformed by the addition of the regulator \eqref{Eq:FRG.2}. The $i$-th operator $\EuScript{O}_i [\Phi]$ has an associated coupling $\bar{g}_i$ which has canonical mass dimension $d_i$. One can define the dimensionless version of $\bar{g}_i$, denoted by $g_i$, using the IR cutoff $k$ as a conversion factor, i.e.,
\begin{equation}
\bar{g}_i = k^{d_i}g_i\,.
\label{Eq:FRG.4}
\end{equation}
Acting with the scale-derivative $\partial_t$ on \eqref{Eq:FRG.4} yields
\begin{equation}
\partial_t\bar{g}_i = d_i k^{d_i}g_i + k^{d_i}\partial_t g_i \equiv d_i k^{d_i}g_i + k^{d_i}\beta_i\,,
\label{Eq:FRG.5}
\end{equation}
with $\beta_i$ denoting the beta function of the coupling $g_i$. Hence, taking the scale-derivative of Eq.\eqref{Eq:FRG.3} and comparing its right-hand side (using Eq.\eqref{Eq:FRG.5}) to the right-hand side of Eq.\eqref{Eq:FRG.1} (expanded in the same basis as \eqref{Eq:FRG.3}), one can read off the beta functions. With such expressions, which are not necessarily perturbative, one can search for FPs. 

A FP $\mathbf{g}^{\ast} \equiv (g_1,\ldots , g_n,\ldots)$ is defined by
\begin{equation}
\beta_i (\mathbf{g}^{\ast}) = 0\,,\quad \forall i\,.
\label{Eq:FRG.6}
\end{equation}
The space of couplings is, in general, infinite-dimensional. As a consequence, one might think that it is necessary to feed the system of beta function with infinitely many initial conditions in order to define the RG flow. If this is the case, the resulting QFT, even featuring a would-be UV FP, is not predictive. However, in general, the existence of a FP imposes severe constraints on the admissible RG trajectories. This is due to the resulting quantum-scale symmetry that emerges at the FP \cite{Hawking:1979ig,Gubitosi:2019ncx}. Analyzing the RG-flow in the neighborhood of the FP, one finds
\begin{equation}
        \beta_{i} (\mathbf{g}) \approx \beta_{i}(\mathbf{g}^{*})+\frac{\partial\beta_{i}}{\partial g_{j}}\Bigg|_{\mathbf{g}^{*}}(g_{j}-g_{j}^{*})\equiv\mathbb{M}_{ij}(g_{j}-g_{j}^{*})\,,
 \label{Eq:FRG.7}
\end{equation}
where only linear terms in $(g-g^\ast)$ were retained. The matrix $\mathbb{M}_{ij}$ is known as the stability matrix. The formal solution of \eqref{Eq:FRG.7} for $g_i$ leads to
\begin{equation}
{g}_{i}(k)= {g}_{i}^{*}+\sum_{I} c_{I}V_{i}^{I}\left(\frac{k}{k_{0}}\right)^{-\theta_{I}}\,,
 \label{Eq:FRG.8}
\end{equation}
where $k_{0}$ is a reference momentum-scale. The numbers $c_I$ are integration constants and the $\theta_I$ are the negative of the eigenvalues of $\mathbb{M}$ and are called critical exponents. Finally, $V^{I}_{i}$ are the corresponding eigenvectors. From the structure in \eqref{Eq:FRG.8}, one sees that for $\theta_I< 0$, the corresponding integration constant $c_I$ must vanish so that the flow of $g_i$ runs to the fixed point at large $k$. This defines an irrelevant or UV-repulsive direction. Conversely, if $\theta_I >0$, the flow runs towards the FP irrespective of the value of $c_I$ and it defines a relevant or UV-attractive direction\footnote{In the situation where the critical exponents are complex, the (ir)relevance of the corresponding direction is defined by the sign of its real part.}. Relevant directions come with free parameters that must be fixed by external conditions, i.e., experiments. As for irrelevant directions, the flow can be described in terms of the above mentioned free parameters and, thus, the corresponding flow becomes a prediction. Hence, predictivity is achieved if the corresponding FP features finitely many relevant directions. Physically relevant theories which are asymptotically safe must feature a predictive FP.   

\subsection{Coarse-Graining in Quantum Gravity}

The Wilsonian renormalization relies on the ordering of momentum scales with respect to a reference cutoff $k$. One can separate the ``fast modes", namely, those with momenta larger than $k$ from the ``slow modes'', which have momentum smaller than $k$. The integration of fast modes together with the progressive reduction of the cutoff scale $k$ defines an iterative procedure that drives the RG flow with respect to $k$. In this procedure, it is essential to have a background structure, i.e., fields are defined on flat space, admitting their description in terms of Fourier modes. Moreover, the hierarchy established between slow and fast modes is possible in a Euclidean setting. 

In quantum gravity, the situation is more subtle. In this case, the geometry of spacetime fluctuates quantum mechanically. Hence, there is no fixed background structure allowing for the definition of a reference scale to discriminate between fast and slow modes. Moreover, the analytic continuation from Euclidean to Lorentzian metrics is challenging, see, e.g., 
\cite{visser2017wickrotategenericcurved,Ambjorn:2001cv}, 
especially if no further constraints are imposed on the configuration space over which the path integral is performed. For this reason, the results discussed in this work are restricted to Euclidean metrics.

The definition of a RG-scale can be achieved by means of the background field method \cite{Abbott:1981ke}. Starting out with the dynamical metric $g_{\mu\nu}$, we split it in a fixed background part $\bar{g}_{\mu\nu}$ and fluctuations $h_{\mu\nu}$ about it, i.e.,
\begin{equation}
g_{\mu\nu} = \bar{g}_{\mu\nu} + h_{\mu\nu}\,.
 \label{Eq:CGQG.1}
\end{equation}
The fluctuations $h_{\mu\nu}$ are not necessarily small. A background Laplacian $-\bar{\nabla}^2$ can be defined, with $\bar{\nabla}$ corresponding to the covariant derivative defined with the connection constructed out of $\bar{g}$. The spectrum of the background Laplacian defines the analogue of momentum in flat space and the dynamical fields can be organized with respect to such a scale. One can think that the introduction of a background metric spoils a defining feature of quantum gravity, namely, background independence. However, such artificial decomposition of the full dynamical field is controlled by Ward identities, ensuring background independence of observables and how generic correlation functions must depend on the background consistently see, e.g., \cite{Labus:2016lkh,Pawlowski:2020qer,Safari:2015dva,Dietz:2015owa,Percacci:2016arh}. Next to that, the linear split between background metric and fluctuation as in Eq.\eqref{Eq:CGQG.1} is just a particular choice. In principle, one is allowed to choose different decompositions. A particularly interesting choice is the so-called exponential parameterization, see, e.g., \cite{Nink:2014yya}, i.e.,
\begin{equation}
g_{\mu\nu} = \bar{g}_{\mu\alpha}{{\left(\mathrm{e}^h\right)}^{\alpha}}_{\nu} = \bar{g}_{\mu\nu} + h_{\mu\nu} + \frac{1}{2} {h_{\mu\alpha}}{h^{\alpha}}_{\nu}+\ldots\,.
 \label{Eq:CGQG.2}
\end{equation}
One remarkable property of the exponential parameterization is the preservation of the metric signature. This means that if the gravitational path integral is performed over Euclidean metrics, one should choose an Euclidean background metric and its signature will not change no matter how large the fluctuations $h_{\mu\nu}$ are, see, e.g., \cite{Demmel_2015,Pawlowski:2020qer,Houthoff:2017oam}. If consistently implemented, the choice of parameterization should not impact observables. 

Besides the conceptual issues associated with coarse-graining in quantum gravity, there is another technicality to be faced: In order to define the graviton propagator, a gauge-fixing term must be introduced. This can be done by the standard Faddeev-Popov procedure \cite{Faddeev1967}. In general, a good criterion to choose the form of the gauge-fixing condition is simplicity. In particular, in the case where functional traces must be taken, as in Eq.~\eqref{Eq:FRG.1}, on a curved background, it is common strategy to search for gauges that make the resulting operator to be traced a minimal operator\footnote{In this case, the highest-order derivative operator contains a Laplacian $-{\nabla}^2$ or powers thereof.}. This is a computationally convenient choice.

In the context of the use of the FRG, there are two issues to be accounted for. The introduction of the regulator might break some symmetries of the original action. Such breaking must be carefully handled by means of modified Ward identities \cite{Labus:2016lkh,Pawlowski:2020qer,Safari:2015dva,Dietz:2015owa,Percacci:2016arh}. This is an intrinsic feature of the FRG and has nothing to do with approximations. At the practical level, truncations must be employed. Working with a given truncation automatically introduces systematic errors that should be minimized. In particular, truncating the dynamics of a given system can introduce spurious unphysical dependences such as on gauge parameters, background fields and field parameterization. On the one hand, depending on the crudeness of the truncation, qualitative features extracted from the FRG might change by variations of those unphysical parameters. On the other hand, such a freedom can be explored by adjusting the parameters to their optimal values, in the sense of speeding up the (apparent) convergence of enlarged truncations. In general, there is no known systematic optimization procedure that can be employed a priori. Exploring simple truncations and refining them can provide indications to the appropriate directions to be optimized. Lastly, the shape function $\mathcal{R}_k$ in the regulator is, up to reasonable properties that a regulator must satisfy, arbitrary. Trustable truncations are those that provide results that are stable against changes in the regulator shape function. Again, in practice, specific choices might be more efficient than others, providing stability of the system sooner \cite{Pawlowski:2015mlf}. See, e.g., the recent work \cite{Riabokon:2025ozw}.

In particular, typical choices for the gauge condition used in the ASQG literature are covariant, linear on the fluctuation $h_{\mu\nu}$ and contain two gauge parameters. In many different setups, the usual scenario is: (i) Both gauge parameters are fixed in order to make the resulting Hessian a minimal operator or (ii) Gauge parameters are left free at the price of the use of the York decompostion (YD) \cite{York:1972sj}. Due to all the systematic uncertainties, if one employs the YD and, after explicit computations, set the gauge parameters to the values used in the minimal case, different quantitative results emerge with respect to the computation using a minimal gauge and not employing the YD. Having good control on this type of scheme dependence is very important to establish the reliability of a given truncation in the context of the FRG. 

In this work, we aim at exploring the robustness of FP structure found within STTs. The strategy relies on the use of the above mentioned spurious dependencies as a reliability-test device. In particular, we shall employ a gauge condition that is not connected to the ``standard'' choice mostly used in the literature by a simple choice of gauge parameters since it involves the dynamical scalar field. This opens up the possibility of testing the FP-candidate structure probing gauge dependence beyond simply gauge-parameter dependence. Additionally, the gauge-fixing employed in this work allows for a direct evaluation of the functional traces without the need of the YD. As previously mentioned, in truncated computations, the use of the YD or not define different schemes that can result in quantitative differences. This can play a pivotal role in deciding whether a given critical exponent is relevant or not if its real part is very close to zero. 

\section{Scalar-Tensor Theories in the Minimal Gauge\label{Sec:STT}}
%
We consider STTs, i.e., theories of modified gravity that, besides the metric, contain a scalar field\footnote{We focus on theories with a single scalar field.}. Most of the developments in the context of STTs within the FRG on a curved background metric were achieved by means of the YD or, alternatively, a flat-space background was employed. In this work, we compute the beta functions in a (maximally symmetric) curved background in this class of theories using a gauge condition that is minimal, i.e., it cancels off all non-minimal terms in the Hessian. A crucial difference with respect to the standard analysis is that the gauge condition depends on the scalar-field fluctuation as well. Hence, this gauge is not a particular case of the standard two-parameters family of covariant gauges usually employed in pure-gravity computations. As a consequence, verifying the existence of a NGFP in this setting corresponds to testing its robustness beyond the exploration of gauge-parameter dependence of the FP. In fact, we are testing the existence of the FP in a new class of gauges. Finding evidence for the existence of the NGFP in this case corresponds to an important step in probing the properties of the NGFP that are gauge independent. 

At this stage, a comment is in order: In an ideal computation where no approximations are employed, one can envision a clear separation between universal and non-universal contributions to the beta functions. This enables the characterization of a fixed-point solution that does not carry spurious gauge-dependent contributions, i.e., changes on the gauge do not affect the FP structure. However, within truncations, it is essential to understand how contaminated the FP solution is with respect to gauge-dependent contributions. If one can really probe the would-be FP within truncations, it is expected that changes on the gauge choice should not engender severe variations on the FP at a qualitative level at least. Hence, finding gauges in which the FP structure is drastically affected might not be taken, immediately, as a hint that the FP candidate is spurious but, instead, that within the underlying approximation, the particular choice of gauge does not optimize (apparent) convergence of the results. If such would-be drastic changes remain as such upon refined truncations, chances that the would-be FP is just a truncation artifact significantly increase. On the other hand, if the main features of the FP are preserved upon significant changes on the gauge choice within a truncation, then a non-trivial stability test is established. This provides evidence for the existence of the FP as a true property of the system instead of a mere truncation artifact. In the following, we discuss the FP-structured obtained for STTs in the minimal gauge and compare our findings with those obtained by the standard gauge choice already reported in the literature.

\subsection{Gravity-Scalar Truncation \label{SubSec:GST}}
%
We shall employ truncations for the Scalar-Tensor EAA\footnote{This is not the full EAA since we still have to introduce gauge-fixing and Faddeev-Popov ghosts terms.} that have the following structure,
\begin{equation}
\Gamma^{\rm ST}_k = \int{\rm d}^dx\sqrt{g}\,\Big(V_k (\Phi)-F_k (\Phi)R + \frac{1}{2}g^{\mu\nu}\partial_\mu \Phi \partial_\nu \Phi \Big)\,.
\label{Eq:SubSec:GST.1}
\end{equation}
The subscript $k$ is introduced in order to emphasize the scale-dependence of the potential $V_k (\Phi)$ and the non-minimal sector $F_k (\Phi)$. The conventions that we use for the geometrical tensors such as the Riemann tensor, Ricci tensor and Ricci scalar are collected in Appendix~\ref{Ap:GeometricO}. For the purposes of this work, we employ the background field method and split the metric as
\begin{equation}
g_{\mu\nu} = \bar{g}_{\mu\nu} + h_{\mu\nu} + \omega\,h_{\mu\alpha}{h^{\alpha}}_{\nu} + \ldots\,.
\label{Eq:SubSec:GST.2}
\end{equation}
This parameterization is interpreted as an interpolating split of the metric that recovers the standard linear split for $\omega = 0$ and the exponential split \eqref{Eq:CGQG.2} for $\omega = 1/2$, at least up to quadratic order in the fluctuation $h_{\mu\nu}$. The choice $\omega = 1$ can also be viewed as the expansion of the metric (up to quadratic order) if the inverse metric is linearly split as $g^{\mu\nu} = \bar{g}^{\mu\nu} - h^{\mu\nu}$. This type of interpolating decomposition has been employed in the literature before, see, e.g., \cite{Kalmykov:1995fd,Kalmykov:1998cv,Gies:2015tca,Ohta:2016npm,Ohta:2016jvw,Goncalves:2017jxq,Ohta:2018sze,DeBrito:2018hur}.

In what follows, we adopt the so-called background approximation (BA) \cite{Reuter:1996cp}. In practice, this is implemented by expanding the full EAA up to second order in the fluctuations. After computing the Hessian that enters \eqref{Eq:FRG.1}, the fluctuations are set to zero. This requires expanding \eqref{Eq:SubSec:GST.1} up to second order in $h_{\mu\nu}$ and on the scalar-field fluctuation $\varphi$ with $\Phi = \bar{\Phi}+\varphi$. The BA is not related to a perturbative expansion around a fixed background but it is an approximation associated with the flow equation \eqref{Eq:FRG.1}. This approximation has known limitations and sophisticated strategies that go beyond the BA are available, see, e.g., \cite{Reichert:2020mja,Pawlowski:2020qer}. Yet, we shall use the BA in the present work for simplicity in order to explore the effects of the minimal gauge to the FP structure. In the next subsection, we describe the form of the gauge-fixing term together with the associated Faddeev-Popov ghosts. 

\subsection{Gauge-fixing and Faddeev-Popov ghosts \label{SubSec:GFFP}}

The ansatz for the EAA \eqref{Eq:SubSec:GST.1} is invariant under general coordinate transformations. In order to be able to compute propagators associated with the fluctuations, a gauge-fixing term is required. This is typically achieved by means of the Faddeev-Popov procedure. 

Inspired by\footnote{More recently, a similar gauge-fixing condition was employed in the context of the Essential Renormalization Group in \cite{Ohta:2025xxo}.} \cite{Shapiro:1995yc,Steinwachs:2011zs}, we choose to work with a gauge-fixing condition that is tuned to cancel the non-minimal terms arising from \eqref{Eq:SubSec:GST.1}. Thus, we define the gauge-fixing action as\footnote{From now on, all quantities with bar are computed with respect to the background fields. Indices are raised/lowered by the background metric.}
\begin{equation}
\Gamma^{\rm gf}_k = \frac{1}{2 \alpha}\int{\rm d}^dx\sqrt{\bar{g}}\,\bar{g}^{\mu\nu}\mathcal{F}_\mu \mathcal{F}_\nu\,,
\label{Eq.SubSec:GFFP.1}
\end{equation}
with
\begin{equation}
\mathcal{F}_\mu = \sqrt{F_k (\bar{\Phi})}\left(\bar{\nabla}^{\alpha}h_{\alpha\mu}-\frac{1+\beta}{d}\bar{\nabla}_\mu h+f_k (\bar{\Phi})\bar{\nabla}_\mu \varphi\right)\,,
\label{Eq.SubSec:GFFP.2}
\end{equation}
where $\alpha$ and $\beta$ are free gauge parameters. The gauge condition \eqref{Eq.SubSec:GFFP.2} reduces to the standard linear covariant gauges, mostly used in the ASQG literature, for $f_k (\bar{\Phi}) = 0$. However, for non-vanishing $f_k (\bar{\Phi})$, a new term, linear on the fluctuation of the scalar field is introduced. This means that the gauge-fixing action contains, besides the usual terms quadratic on $h_{\mu\nu}$, terms quadratic on $\varphi$ and mixed $(h,\varphi)$ terms. In principle, the function $f_k (\bar{\Phi})$ can be an arbitrary function of the background field $\bar{\Phi}$. However, we choose it in order to cancel non-minimal terms arising from the expansion of $\Gamma^{\rm ST}_k$ in powers of the fluctuations. Explicitly, the general form of \eqref{Eq.SubSec:GFFP.1} is\footnote{We shall use the shorthand notation $\int_x \equiv \int {\mathrm{d}}^dx\sqrt{\bar{g}}$ when convenient.}
\begin{eqnarray}
\Gamma^{\rm gf}_k 
&=& \frac{1}{2\alpha}\int_x\,F_k (\bar{\Phi})\Bigg[\bar{\nabla}^{\alpha}h_{\alpha\mu}\bar{\nabla}^{\lambda}{h_{\lambda}}^{\mu}-2\frac{1+\beta}{d}\bar{\nabla}^{\alpha}h_{\alpha\mu}\bar{\nabla}^\mu h \nonumber\\
&+& 2(\bar{\nabla}^{\alpha}h_{\alpha\mu}) f_k (\bar{\Phi})\bar{\nabla}^\mu\varphi
+\Bigg(\frac{1+\beta}{d}\Bigg)^2\bar{\nabla}_\mu h \bar{\nabla}^\mu h\nonumber\\
&-&2\frac{1+\beta}{d}(\bar{\nabla}_\mu h)f_k (\bar{\Phi})\bar{\nabla}^\mu\varphi+(f_k(\bar{\Phi}))^2\bar{\nabla}_\mu\varphi \bar{\nabla}^\mu\varphi\Bigg]\,.\nonumber\\
\label{Eq.SubSec:GFFP.3}
\end{eqnarray}
In order to cancel out all the non-minimal terms, we choose
\begin{equation}
\alpha = 1\,, \quad \beta = \frac{d}{2}-1\,, \quad \mathrm{and} \quad f_k (\bar{\Phi})=- \frac{F^\prime_k (\bar{\Phi})}{F_k (\bar{\Phi})}\,.
\label{Eq.SubSec:GFFP.4}
\end{equation}
This is a Feynman-de Donder-like gauge which we shall call, minimal Feynman-de Donder-Scalar (mFDS) gauge for convenience. By definition, the Faddeev-Popov ghosts action is
\begin{eqnarray}
\Gamma^{\rm gh}_k  &=& \int_x \sqrt{F_k(\bar{\Phi})}\,\bar{C}_\mu \left\{ \bar{\nabla}^\alpha \left(g_{\mu\nu}\nabla_\alpha 
+ g_{\alpha\nu}\nabla_\mu\right)\right.\nonumber\\
&-&\left.\frac{1+\beta}{d}\bar{\nabla}_\mu\left[\bar{g}^{\alpha\beta}\left(g_{\beta\nu} \nabla_{\alpha} + g_{\alpha\nu}\nabla_\beta \right)\right]\right.\nonumber\\
&+&\left.f_k (\bar{\Phi})\bar{\nabla}_\mu \left[g^{\lambda\alpha}(\nabla_\alpha \Phi) g_{\lambda \nu}\right]\right.\nonumber\\
&+&\left.f_k (\bar{\Phi})\left[g^{\lambda\alpha}(\nabla_\alpha \Phi) g_{\lambda\nu}\right]\bar{\nabla}_\mu\right\} C^\nu\,,
\label{Eq.SubSec:GFFP.5}
\end{eqnarray}
with $(\bar{C}_\mu,C^\nu)$ being the Faddeev-Popov anti-ghost and ghost fields. Some comments are in order: Action \eqref{Eq.SubSec:GFFP.5} involves the full metric $g_{\mu\nu}$ and its inverse and the full scalar field $\Phi$. Hence, in principle, there are infinitely many vertices involving ghosts, metric and scalar fluctuations. Since we are working within the BA and the ghost action is bilinear in the ghosts, we can replace the full metric and (full) scalar field by their background counterparts in \eqref{Eq.SubSec:GFFP.5}. The first two terms of \eqref{Eq.SubSec:GFFP.5} correspond to the standard ghost action for the usual linear covariant gauges. The mFDS gauge introduces the last two terms. The global factor associated with $F_k (\bar{\Phi})$ is a function of the background scalar field and its presence (or absence) is associated with the choice of measure of the path integral. For our purposes, we can perform our analysis in the presence or absence of such a factor as in, e.g., \cite{Eichhorn:2020sbo,Narain:2009fy}. Simplifying the ghost action due to the BA and enforcing condition \eqref{Eq.SubSec:GFFP.4} leads to 
\begin{eqnarray}
\Gamma^{\rm gh}_k &=& \int_x \sqrt{F_k(\bar{\Phi})}\,\bar{C}_\mu \left[ \delta^\mu_\nu \bar{\nabla}^2 
+f_k (\bar{\Phi})(\bar{\nabla}^\mu \bar{\nabla}_\nu \bar{\Phi}) \right.\nonumber\\
&+&\left.f_k (\bar{\Phi})(\bar{\nabla}_\nu \bar{\Phi})\bar{\nabla}^\mu +{\bar{R}^{\mu}\,}_\nu\right]C^\nu\,.
\label{Eq.SubSec:GFFP.6}
\end{eqnarray}
For the purposes of this work, we will take the following simplifying choices for the background fields: The background scalar field is chosen to be constant\footnote{For the sake of clarity, we denote the constant scalar field with a subscript namely $\bar{\Phi}_0$. However, unless otherwise stated, we will simply call the constant background scalar field as $\bar{\Phi}$.} and the background geometry is taken to be maximally symmetric, i.e., 
\begin{eqnarray}
\bar{\Phi} &=& \bar{\Phi}_0\,, \quad \bar{R}_{\mu\nu\alpha\beta} = \frac{\bar{R}}{d(d-1)}(\bar{g}_{\mu\alpha}\bar{g}_{\nu\beta} - \bar{g}_{\mu\beta}\bar{g}_{\nu\alpha})\,, \nonumber\\
\bar{R}_{\mu\nu} &=& \frac{\bar{R}}{d}\bar{g}_{\mu\nu}\,,  
\label{Eq:Setup12}
\end{eqnarray}
with $\bar{R}$ being constant. This particular choice for the background fields considerably simplifies the ghost action since those terms that depend on the derivative of the background scalar fields vanish. We are now ready to compute the Hessian $\Gamma^{(2)}_k$ of the complete EAA,
\begin{equation}
    \Gamma_k = \Gamma^{\rm ST}_k + \Gamma^{\rm gf}_k + \Gamma^{\rm gh}_k\,.
\end{equation}
and its regularized version.

\section{Regularized Hessian and Traces}
\label{Sect:Hessian}

\subsection{Hessian and its Regularization}

In order to compute the Hessian, one expands the action\footnote{We employ the collective notation $\chi$ for the fluctuation fields and $\bar{\chi}$ for their background counterparts.} $\Gamma_k [\chi,\bar{\chi}]$ up to second order in the fluctuation fields $\chi$, see Appendix~\ref{Ap:Fluctuations}. Here, we report directly the Hessian and its components are represented by
\begin{equation}
\Gamma^{\chi \chi}_{k,IJ}(x,y)=\frac{1}{\sqrt{\bar{g}(x)}}\frac{1}{\sqrt{\bar{g}(y)}}\frac{\delta^2\Gamma_k}{\delta \chi_I (x) \delta \chi_J (y)}\Bigg|_{\chi = 0}\,.
\label{Eq:Hessian.0}
\end{equation} 
We introduce the traceless and trace projectors as follows,
\begin{equation}
{\Pi}^{\rm TL}_{\mu\nu,\alpha\beta} = \frac{1}{2}\left(\bar{g}_{\mu\alpha}\bar{g}_{\nu\beta} + \bar{g}_{\mu\beta}\bar{g}_{\nu\alpha}\right)-\frac{1}{d}\bar{g}_{\mu\nu}\bar{g}_{\alpha\beta}\,, 
\label{Eq:Hessian.1}
\end{equation}
and
\begin{equation}
{\Pi}^{\rm T}_{\mu\nu,\alpha\beta} = \frac{1}{d}\bar{g}_{\mu\nu}\bar{g}_{\alpha\beta}\,.
\label{Eq:Hessian.1.1}
\end{equation}
Being projectors, they satisfy the following relations
\begin{eqnarray}
\left[{\Pi}^{\rm TL}\right]\cdot \left[{\Pi}^{\rm TL}\right] &=& \left[{\Pi}^{\rm TL}\right]\,,\quad \left[{\Pi}^{\rm T}\right] \cdot \left[{\Pi}^{\rm T}\right] = \left[{\Pi}^{\rm T}\right]\,,\nonumber\\
\left[{\Pi}^{\rm TL}\right] \cdot \left[{\Pi}^{\rm T}\right] &=& \left[{\Pi}^{\rm T}\right] \cdot \left[{\Pi}^{\rm TL}\right] = 0 \,,
\label{Eq:Hessian.1.2}
\end{eqnarray}
and
\begin{equation}
\left[{\Pi}^{\rm TL}\right]+\left[{\Pi}^{\rm T}\right] = \mathbf{1}_{(2)}\,.
\label{Eq:Hessian.1.3}
\end{equation}
The symbol $\mathbf{1}_{(2)}$ represents the identity in the space of symmetric rank-2 tensors. The product of projectors represented above assumes the correct contraction of indices and we omit it by employing a matrix-like notation.

Explicitly, the graviton-graviton component of the Hessian reads
\begin{eqnarray}
\Gamma^{hh}_k &=& \frac{1}{2}\left[F_k(\bar{\Phi})\Delta + \zeta(\omega)V_k (\bar{\Phi})+F_k (\bar{\Phi})\bar{R}\,\zeta_1 (d,\omega)\right]\,\Pi^{\rm TL} \nonumber\\
&-& \frac{d-2}{4}\left[F_k (\bar{\Phi})\Delta - \frac{d+2\zeta(\omega)}{d-2}V_k (\bar{\Phi})\right.\nonumber\\
&-&\left.F_k (\bar{\Phi})\frac{\bar{R}}{d-2}\zeta_2 (d,\omega)\right]\,\Pi^{\rm T}\,,
\label{Eq:Hessian.2}
\end{eqnarray}
with
\begin{equation}
    \zeta_1 (d,\omega) = \frac{2+\zeta(\omega)(2-d)(d-1)}{d(d-1)}\,,
\end{equation}
and
\begin{eqnarray}
\zeta_2 (d,\omega) &=& -\frac{1}{d(d-1)}(d^3-d^2(5-2\zeta(\omega))-d(6\zeta(\omega)-8)\nonumber\\
&+&4(\zeta(\omega)-1))\,,
\label{Eq:Hessian.3}
\end{eqnarray}
where $\zeta(\omega) = 2\omega - 1$ and $\Delta \equiv -\bar{\nabla}^2$. 
The purely-scalar component of the Hessian is given by
\begin{equation}
\Gamma^{\varphi\varphi}_k = \left[\frac{(F^\prime_k (\bar{\Phi}))^2}{F_k (\bar{\Phi})}+1\right]\Delta - F^{\prime\prime}_k (\bar{\Phi})\bar{R} + V^{\prime\prime}_k (\bar{\Phi})\,.
\label{Eq:Hessian.5}
\end{equation}
The mixed graviton-scalar components are
\begin{eqnarray}
\Gamma^{h\varphi}_k &=& \left[-\frac{1}{2} F^\prime (\bar{\Phi})\Delta + \left(\frac{2-d}{2d}\right)\bar{R} F^\prime (\bar{\Phi}) + \frac{1}{2} V^\prime_k (\bar{\Phi})\right]\mathbf{1}_{(1)}\nonumber\\
&=& \Gamma^{\varphi h}_k\,. 
\label{Eq:Hessian.6}
\end{eqnarray}
with $\mathbf{1}_{(1)}$ being the identity in the space of spin 1 fields. Finally, the components of the Hessian associated with the ghost-sector are,
\begin{equation}
\Gamma^{c\bar{c}}_k  =-\Gamma^{\bar{c}c}_k = -\sqrt{F_k (\bar{\Phi})}\left(\Delta - \frac{\bar{R}}{d}\right)\mathbf{1}_{(1)}\,.
\label{Eq:Hessian.7}
\end{equation}
The complete Hessian in matrix notation is written as
\begin{equation}
\Gamma^{(2)}_k = \begin{pmatrix}
\Gamma^{hh}_k & \Gamma^{h\varphi}_k & 0 & 0\\ 
\Gamma^{\varphi h}_k & \Gamma^{\varphi\varphi}_k & 0 & 0 \\
0 & 0 & 0 & \Gamma^{c\bar{c}}_k \\
0 & 0 & \Gamma^{\bar{c}c}_k & 0 \\
\end{pmatrix}\,.
\label{Eq:Hessian.8}
\end{equation} 
Next to that, one has to regularize the Hessian $\Gamma^{(2)}_k$. This amounts to add a matrix $\mathbb{R}_k$ with the structure
\begin{equation}
\mathbb{R}_k = \begin{pmatrix}
\mathbb{R}^{hh}_k & \mathbb{R}^{h\varphi}_k & 0 & 0\\ 
\mathbb{R}^{\varphi h}_k & \mathbb{R}^{\varphi\varphi}_k & 0 & 0 \\
0 & 0 & 0 & \mathbb{R}^{c\bar{c}}_k \\
0 & 0 & \mathbb{R}^{\bar{c}c}_k & 0 \\
\end{pmatrix}\,,
\label{Eq:Hessian.9}
\end{equation} 
to $\Gamma^{(2)}_k$. We opt for a regularization where all Laplacians are exchanged by their regularized counterpart. Explicitly, the regulators are given by
\begin{equation}
\mathbb{R}^{hh}_k = \frac{1}{2}\Big(F_k(\bar{\Phi})R_k (\Delta)\Big)\,\Pi^{\rm TL} 
- \frac{d-2}{4}\Big(F_k (\bar{\Phi})R_k (\Delta) \Big)\,\Pi^{\rm T}\,,
\label{Eq:Hessian.11}
\end{equation}
\begin{equation}
\mathbb{R}^{\varphi\varphi}_k = \Bigg[\frac{(F^\prime_k (\bar{\Phi}))^2}{F_k (\bar{\Phi})}+ 1\Bigg]\,R_k (\Delta)  \,.
\label{Eq:Hessian.12}
\end{equation}
\begin{equation}
\mathbb{R}^{h\varphi}_k = \mathbb{R}^{\varphi h}_k = \left[-\frac{1}{2} F^\prime (\bar{\Phi})\,R_k (\Delta)\right]\mathbf{1}_{(1)}\,,
\label{Eq:Hessian.13}
\end{equation}
and
\begin{equation}
\mathbb{R}^{c\bar{c}}_k  = -\mathbb{R}^{\bar{c}c}_k =  \left[-\sqrt{F_k (\bar{\Phi})}\,R_k (\Delta) \right]\mathbf{1}_{(1)}\,.
\label{Eq:Hessian.14}
\end{equation}
Finally, in order to completely specify the regulator, we choose the Litim \cite{Litim:2001up} shape function $R_k (\Delta)$,
\begin{equation}
R_k (\Delta) = (k^2 - \Delta)\theta (k^2 - \Delta)\,,
\label{Eq:Hessian.15}
\end{equation}
with $\theta (x)$ being the step function. At this point, it is important to stress that the regulators that have been chosen above depend, in general, on the background scalar field $\bar{\Phi}$. Moreover, they also contain scale-dependent couplings that appear in the functions $F_k (\bar{\Phi})$. Having a good control on how such choices affect the results is not an easy task and one should be careful with definite conclusions, see, e.g., \cite{Bridle:2013sra}. It is well-known that an uncontrolled dependence on the background in the regulator can lead to potential problems already in simple systems.

The regularized Hessian has the following structure,
\begin{equation}
\Gamma^{(2)}_k +\mathbb{R}_k = \begin{pmatrix}
\mathbb{M}_{\rm BB} &  0\\ 
0 & \mathbb{M}_{\rm GG} \\
\end{pmatrix}\,,
\label{Eq:Hessian.16}
\end{equation} 
with $\mathbb{M}_{\rm BB}$ standing for the matrix which contains all the contributions from bosonic fields and $\mathbb{M}_{\rm GG}$ contains all the contributions from the ghosts (Grassmannian fields). Since there is no mixing between those sectors we can deal with each of them separately. In the next Subsection, we describe the structure of the traces that appear in the flow equation for those different sectors.

\subsection{Taking Traces}

The key element in the flow equation is the evaluation of a regularized trace. However, a crucial step before the evaluation of the corresponding traces is the inversion of the regularized Hessian. One can follow several different routes at this stage. Firstly, we consider the bosonic sector which is expressed as 
\begin{equation}
\Gamma^{(2)}_{{(\rm BB)},k} (P_k) = \begin{pmatrix}
\Gamma^{hh}_k (P_k) & \Gamma^{h\varphi}_k (P_k) \\ 
\Gamma^{\varphi h}_k (P_k) & \Gamma^{\varphi\varphi}_k (P_k) \\
\end{pmatrix}\,,
\label{Eq:Trace.1}
\end{equation} 
where $\Gamma^{hh}_k (P_k) \equiv \Gamma^{hh}_k +\mathbb{R}^{hh}_k$ and so on, i.e., the elements of \eqref{Eq:Trace.1} are the regularized bosonic components of the Hessian. It is convenient to write
\begin{equation}
\Gamma^{hh}_k (P_k) = \Gamma^{hh}_{{\rm TL},k} (P_k)\, \Pi^{\rm TL} + \Gamma^{hh}_{{\rm T},k} (P_k)\, \Pi^{\rm T}\,.
\label{Eq:Trace.2}
\end{equation}
In the flow equation \eqref{Eq:FRG.1}, one needs to invert \eqref{Eq:Trace.2}. This can be achieved by standard algebraic manipulations leading to the following inverse matrix,
\begin{equation}
\left[\Gamma^{(2)}_{{(\rm BB)},k} (P_k)\right]^{-1} = \begin{pmatrix}
\mathbb{G}^{(hh)}_{k} & \mathbb{G}^{(h\varphi)}_{k} \\ 
\mathbb{G}^{(\varphi h)}_{k} & \mathbb{G}^{(\varphi \varphi)}_{k} \\
\end{pmatrix}\,,
\label{Eq:Trace.3}
\end{equation} 
with
\begin{eqnarray}
\mathbb{G}^{(hh)}_{k} &\equiv& \frac{1}{\Gamma^{hh}_{{\rm TL},k} (P_k)}\, \Pi^{\rm TL} \nonumber\\
&+& \frac{\Gamma^{\varphi\varphi}_k (P_k)}{\Gamma^{hh}_{{\rm T},k} (P_k) \Gamma^{\varphi\varphi}_k (P_k) - d [\Gamma^{\varphi h}_k (P_k)]^2}\, \Pi^{\rm T}\,,
\end{eqnarray}
\begin{eqnarray}
\mathbb{G}^{(h\varphi)}_k &\equiv& - \frac{\Gamma^{h\varphi}_k (P_k)}{\Gamma^{hh}_{{\rm T},k} (P_k) \Gamma^{\varphi\varphi}_k (P_k) - d [\Gamma^{\varphi h}_k (P_k)]^2}\,\mathbf{1}_{(1)}\nonumber\\
&\equiv& \mathbb{G}^{(\varphi h)}_k\,,
\end{eqnarray}
\begin{eqnarray}
\mathbb{G}^{(\varphi \varphi)}_k \equiv  \frac{\Gamma^{hh}_{{\rm T},k} (P_k)}{\Gamma^{hh}_{{\rm T},k} (P_k) \Gamma^{\varphi\varphi}_k (P_k) - d [\Gamma^{\varphi h}_k (P_k)]^2}\,.
\label{Eq:Trace.4}
\end{eqnarray}
Next to that, we have to evaluate the trace of the matrix resulting from the product of \eqref{Eq:Trace.3} and the scale-derivative $\partial_t $ of the bosonic sector of \eqref{Eq:Hessian.9}. Such an operation reduces to the following traces, 
\begin{widetext}
\begin{equation}
\EuScript{T}_{(2)} = {\rm Tr}\left[\frac{\partial_t\mathbb{R}^{hh}_{{\rm TL,k}}}{\Gamma^{hh}_{{\rm TL},k} (P_k)}\, \Pi^{\rm TL} + \left(\frac{\Gamma^{\varphi\varphi}_k (P_k)\partial_t\mathbb{R}^{hh}_{{\rm T,k}}}{\Gamma^{hh}_{{\rm T},k} (P_k) \Gamma^{\varphi\varphi}_k (P_k) - d [\Gamma^{\varphi h}_k (P_k)]^2}-d\frac{\Gamma^{h\varphi}_k (P_k) \partial_t\mathbb{R}^{h\varphi}_{\rm{k}}}{\Gamma^{hh}_{{\rm T},k} (P_k) \Gamma^{\varphi\varphi}_k (P_k) - d [\Gamma^{\varphi h}_k (P_k)]^2}\right)\, \Pi^{\rm T}\right]\,,
\label{Eq:Trace.5}
\end{equation}
and
\begin{equation}
\EuScript{T}_{(0)} = {\rm Tr}\left[\frac{\Gamma^{hh}_{{\rm T},k} (P_k) \partial_t \mathbb{R}^{\varphi\varphi}_{k}}{\Gamma^{hh}_{{\rm T},k} (P_k) \Gamma^{\varphi\varphi}_k (P_k) - d [\Gamma^{\varphi h}_k (P_k)]^2}-d\frac{\Gamma^{\varphi h}_k (P_k)\partial_t \mathbb{R}^{h\varphi}_k}{\Gamma^{hh}_{{\rm T},k} (P_k) \Gamma^{\varphi\varphi}_k (P_k) - d [\Gamma^{\varphi h}_k (P_k)]^2}\right]\,.
\label{Eq:Trace.6}
\end{equation}
\end{widetext}
The traces that contribute to the bosonic sector of the flow equation are collectd in \eqref{Eq:Trace.5} and \eqref{Eq:Trace.6}. Now, we turn to the ghost sector. Following the same procedure as the one just described for the bosonic sector, the resulting trace to be computed arising from the Grassmannian sector of the Hessian is
\begin{equation}
\EuScript{T}_{(1)} = {\rm Tr}\left[\frac{\partial_t \mathbb{R}^{\bar{c} c}_k }{\Gamma^{\bar{c}{c}}_k (P_k)}\mathbf{1}_{(1)}\right]\,.
\label{Eq:Trace.7}
\end{equation}
Collecting the contributions \eqref{Eq:Trace.5}, \eqref{Eq:Trace.6} and \eqref{Eq:Trace.7}, the flow equation is expressed as
\begin{equation}
\partial_t \Gamma_k = \frac{1}{2}\EuScript{T}_{(2)} + \frac{1}{2}\EuScript{T}_{(0)} - \EuScript{T}_{(1)} \,,
\label{Eq:Trace.8}
\end{equation}
where the minus sign in front of $\EuScript{T}_{(1)}$ is due to the Grassmannian nature of this contribution. In \eqref{Eq:Trace.8}, $\Gamma_k$, in the BA, is given by
\begin{equation}
\Gamma_k \equiv \Gamma^{\rm ST}_k [\bar{g},\bar{\Phi}] \,.
\label{Eq:Trace.9}
\end{equation}
The next step of the present analysis consists in computing the traces explicitly and, by a suitable projection rule, reading off the beta functions of the underlying couplings. In this work, the computation of the traces is performed by means of heat kernel techniques. In the next section, we collect our findings for fixed points in $d=4$ and compare them with existing results in the literature. Before moving to the explicit computations, since we are concerned with the fixed points associated with the dimensionless counterparts of the couplings and we express the quantities involved the the flow equation in units of $k$, i.e.,
\begin{equation}
\Phi = k^{(d-2)/2}\tilde{\Phi}\,,
\label{Eq:Trace.10}
\end{equation}
where the tilde is used to emphasize that such quantity is dimensionless. Therefore
\begin{eqnarray}
V_k (\Phi) = k^d \tilde{V}_k (\tilde{\Phi})\,,\quad {\rm and} \quad
F_k (\Phi) = k^{d-2} \tilde{F}_k (\tilde{\Phi})\,.
\label{Eq:Trace.11}
\end{eqnarray}
For the purposes of this work our analysis is limited to polynomial choices for the functions $V_k (\Phi)$ and $F_k (\Phi)$. Our truncations are established by the degree of each polynomial. In particular,
\begin{equation}
V_k (\Phi) = \sum^a_{n=0} \lambda_{2n}\Phi^{2n}\,,\quad {\rm and} \quad F_k (\Phi) = \sum^{b}_{n=0}\xi_{2n} \Phi^{2n}\,,
\label{Eq:Trace.12}
\end{equation}
and such a truncation is labeled as $(a,b)$. In order to read off the beta functions from the flow equations, suitable projection rules are necessary. In particular, we define the following projection operator
\begin{equation}
    \mathcal{P}_{2n}[f]=\frac{1}{(2n!)}\frac{\partial^{2n}f(\tilde{\Phi})}{\partial \tilde{\Phi}^{2n}}\Bigg|_{\tilde{\Phi}=0}.
\end{equation}
Therefore, we obtain the running of each one of the beta functions in (\ref{Eq:Trace.12}) by making the following projection
\begin{equation}
    \partial_{t}\tilde{\lambda}_{2n}=\mathcal{P}_{2n}[\partial_{t}\tilde{V}_{k}],\;\; \partial_{t}\tilde{\xi}_{2n}=\mathcal{P}_{2n}[\partial_{t}\tilde{F}_{k}]\,.
\end{equation}

Before we move to the FP-analysos, a summary of our setup is in order: The gauge condition defined by Eq.\eqref{Eq.SubSec:GFFP.2} supplied by the choices \eqref{Eq.SubSec:GFFP.4} allows for the elimination of all non-minimal terms in the Hessian. This is in contrast with the standard gauge-fixing condition, see, e.g., \cite{Narain:2009fy}, where a YD is employed in order to eliminate the remaining non-minimal terms. Hence, in the present setup, besides testing the existence of FP candidates in a different class of gauges that is not deformable to the standard linear covariant gauges mostly employed in the literature, we can also use standard heat-kernel techniques directly. As discussed before, the use of the YD or not lead to a change of computational scheme that produce quantitative differences. Even if they are small, some of the quantities we are interested in this work (such as the real part of some critical exponents associated with some scalar-field couplings) turn out to be very small. Therefore, small changes arising from different computational schemes can, in principle, flip the sign of such quantities of interest and this changes, for instance, the number of free parameters of the underlying theory. Such a comment illustrates that it is crucial to understand how different choices of computational schemes can concretely affect the FP structure qualitatively. Thus, the present choice of gauge allows for testing a different class of gauge condition and of computational scheme in which the YD is not employed.

\section{Fixed-point Analysis in Four dimensions}
\label{Sect:FP4d}

Our main interest is to search for fixed-point candidates that can render STTs UV complete in four dimensions. Since we are employing a gauge-fixing condition that does not belong to the standard two-parameters family of gauges usually employed in the literature of ASQG, our first concern is to establish evidence for the existence of such a FP in the present case. We limit ourselves to the class of Gaussian Matter Fixed Points (GMFP), see \cite{Narain:2009fy}. The eventual existence of FPs that fall outside this class is left for future investigation. This class of FPs develops non-trivial values for the gravitational couplings whereas those associated with the scalar-field sector vanish (thus Gaussian). 

The simplest truncation involves field-independent contributions to $V(\Phi)$ and $F(\Phi)$. In particular, $\tilde{V}_k (\tilde{\Phi}) = \tilde{\lambda}_0$ and $\tilde{F}_k (\tilde{\Phi}) = \tilde{\xi}_0$. Those couplings can be written in terms of the dimensionless Newton and cosmological constants, $\tilde{G}_k$ and $\tilde{\Lambda}_k$, respectively, as
\begin{equation}
\tilde{\lambda}_0 = \frac{2\tilde{\Lambda}_k}{16 \pi \tilde{G}_k}\,,\quad {\rm and} \quad \tilde{\xi}_0 = \frac{1}{16 \pi \tilde{G}_k}\,.
\label{Eq:FP4d.1}
\end{equation}
The EAA reduces to the Einstein-Hilbert truncation together with a kinetic term for the scalar field quantized in the mFDS gauge. By setting the interpolating parameter to $\omega = 0$, i.e., by employing the linear split of the metric, a non-trivial fixed point for the gravitational couplings is found,\footnote{For the sake of simplicity, we will omit the tilde over dimensionless couplings from now on unless there is a potential source of confusion.}
\begin{equation}
{\Lambda}^\ast = 0.182\,,\quad {\rm and} \quad {G}^\ast = 0.724\,\quad (\omega = 0)\,,
\label{Eq:FP4d.2}
\end{equation}
with critical exponents 
\begin{equation}
\theta_{1,2} = 1.829 \pm 2.759\,i\,. 
\label{Eq:FP4d.2.1}
\end{equation}
This result can be directly compared with the one reported in \cite{Narain:2009fy} where the computation was performed using a linear parameterization of the metric together with the York decomposition. Results for Landau-de Donder gauge ($\alpha = 0$ and $\beta=1$ in the standard two-parameters linear covariant gauges) are: $(\Lambda^\ast,{G}^\ast)_{\rm NP}= (0.181,0.837)$ and the relevant critical exponents are $(\theta_{1,2})_{\rm NP} = 2.143\pm 2.879\,i$, where we use the label NP to refer to the authors of \cite{Narain:2009fy}. The real parts of the critical exponents obtained in the mFDS gauge are 15\% off with respect to the ones reported in \cite{Narain:2009fy}. Considering that the computations were performed under significantly different schemes, e.g., different gauges and use of YD, and the crudeness of the truncation, the results are rather close.

Regarding the exponential parameterization $\omega = 1/2$, the FP reads
\begin{equation}
{\Lambda}^\ast = 0.189\,,\quad {\rm and} \quad {G}^\ast = 1.038\,\quad (\omega = 1/2)\,,
\label{Eq:FP4d.3}
\end{equation}
with critical exponents 
\begin{equation}
\theta_{1,2} = 1.642 \pm 6.362\,i\,. 
\label{Eq:FP4d.3.1}
\end{equation}
Our results are about 37\% off with respect to the ones reported in \cite{Ohta:2021bkc} in the (standard) Landau-de Donder gauge using the York decomposition. 

We single out the value $\omega = 1$ as it represents a linear split of the inverse metric. In this case, the  FP reads
\begin{equation}
{\Lambda}^\ast = 0.117\,,\quad {\rm and} \quad {G}^\ast = 1.001\,\quad (\omega = 1)\,,
\label{Eq:FP4d.4}
\end{equation}
with critical exponents 
\begin{equation}
\theta_{1,2} = 1.908 \pm 6.793\,i\,. 
\label{Eq:FP4d.4.1}
\end{equation}
To our knowledge, the corresponding results within this parameterization were not reported anywhere. 

In the three parameterizations we singled out, the critical exponents are complex and the FP values are quantitatively close. Since they were obtained with a gauge condition that is not deformable to the standard linear covariant one by simply changing gauge parameters, one sees that those results probe not just gauge-parameter but gauge-fixing dependence when compared to the other results in the literature. 

From now on, we label our truncations by pairs $(a,b)$ which corresponds to the upper limits of the sums in \eqref{Eq:Trace.12}. Most of our attention will be devoted to the linear $(\omega=0)$, exponential $(\omega=1/2)$ and linear-inverse $(\omega=1)$ parameterizations. Our findings reveal common features in all cases (we have analyzed truncations up to $(4,4)$):
\begin{itemize}
\item If we enlarge the truncation in the ${V}_k ({\Phi})$-direction, i.e., we keep $b=1$ and increase the value of $a$, a FP candidate featuring two relevant directions persists;
\item Adding terms in the ${F}_k ({\Phi})$-direction does not change this picture qualitatively. In all cases, i.e., in the linear, exponential and linear-inverse parameterizations, a FP candidate is found at each truncation order with two relevant directions.
\end{itemize}
The values of the FPs are given by \eqref{Eq:FP4d.2},  \eqref{Eq:FP4d.3} and \eqref{Eq:FP4d.4} together with all matter couplings being zero, i.e., ${\lambda}^\ast_{2n} = 0$ and ${\xi}^\ast_{2n} = 0$ for $n\ge1$. As for the critical exponents, we collect our results in Tables \ref{Table1}, \ref{Table2} and \ref{Table3}.
\begin{table*}[t!]
\centering
 \begin{tabular}{|| c c c c c c||} 
 \hline
 $(a,b)$ & $\theta_{1,2}$ & $\theta_{3,4}$ & $\theta_{5}$ & $\theta_{6}$ & $\theta_{7}$ \\ [0.5ex] 
 \hline\hline
 (4,1) & $1.829 \pm 2.759\,i$ & $-0.170 \pm 2.759\,i$ & -3.823 & -5.823 & -7.823 \\
 \hline
 $(a,b)$ & $\theta_{1,2}$ & $\theta_{3,4}$ & $\theta_{5,6}$  & $\theta_{7}$ & $\theta_{8}$\\ [0.5ex]
 \hline\hline
 (4,2) & $1.829 \pm 2.759\,i$ &  $-0.170 \pm 2.759\,i$ & $-2.170 \pm 2.759\,i$ & -5.823 & -7.823\\
 \hline
 $(a,b)$ & $\theta_{1,2}$  & $\theta_{3,4}$ & $\theta_{5,6}$  & $\theta_{7,8}$ & $\theta_{9}$ \\ [0.5ex]
 \hline\hline
 (4,3) & $1.829 \pm 2.759\,i$ &  $-0.170 \pm 2.759\,i$ & $-2.170 \pm 2.759\,i$ & $-4.170 \pm 2.759\,i$ & -7.823 \\
\hline
  $(a,b)$ & $\theta_{1,2}$  & $\theta_{3,4}$  & $\theta_{5,6}$  & $\theta_{7,8}$  & $\theta_{9,10}$ \\ [0.5ex]
  \hline\hline
 (4,4) & $1.829 \pm 2.759\,i$ &  $-0.170 \pm 2.759\,i$ & $-2.170 \pm 2.759\,i$ & $-4.170 \pm 2.759\,i$ & $-6.170 \pm 2.759\,i $\\
 \hline
 \end{tabular}
 \caption{Critical exponents in the linear parametrization.  For each truncation, a FP with two relevant directions was found in $d=4$. We refine the truncation both in $V$- and $F$-directions.}
 \label{Table1}
\end{table*}
\begin{table*}[t!]
\centering
 \begin{tabular}{|| c c c c c c||} 
 \hline
 $(a,b)$ & $\theta_{1,2}$ & $\theta_{3,4}$ & $\theta_{5}$ & $\theta_{6}$ & $\theta_{7}$ \\ [0.5ex] 
 \hline\hline
 (4,1) & $1.642 \pm 6.362\,i$ & $-0.357  \pm 6.362\,i$ & -7.489 & -9.489 & -11.489 \\
 \hline
 $(a,b)$ & $\theta_{1,2}$ & $\theta_{3,4}$ & $\theta_{5,6}$  & $\theta_{7}$ & $\theta_{8}$\\ [0.5ex]
 \hline\hline
 (4,2) & $1.642 \pm 6.362\,i$ &  $-0.357 \pm 6.362\,i$ & $-2.357 \pm 6.362\,i$ & -9.489 & -11.489\\
 \hline
 $(a,b)$ & $\theta_{1,2}$  & $\theta_{3,4}$ & $\theta_{5,6}$  & $\theta_{7,8}$ & $\theta_{9}$ \\ [0.5ex]
 \hline\hline
 (4,3) & $1.642 \pm 6.362\,i$ &  $-0.357 \pm 6.362\,i$ & $-2.357 \pm 6.362\,i$ & $-4.357 \pm 6.362\,i$ & -11.489 \\
\hline
  $(a,b)$ & $\theta_{1,2}$  & $\theta_{3,4}$  & $\theta_{5,6}$  & $\theta_{7,8}$  & $\theta_{9,10}$ \\ [0.5ex]
  \hline\hline
 (4,4) & $1.642 \pm 6.362\,i$ &  $-0.357 \pm 6.362\,i$ & $-2.357 \pm 6.362\,i$ & $-4.357 \pm 6.362\,i$ & $-6.357\pm6.362\,i$\\
 \hline
 \end{tabular}
 \caption{Critical exponents in the exponential parametrization.  For each truncation, a FP with two relevant directions was found in $d=4$. We refine the truncation both in $V$- and $F$-directions.}
 \label{Table2}
\end{table*}
\begin{table*}[t!]
\centering
 \begin{tabular}{|| c c c c c c||} 
 \hline
 $(a,b)$ & $\theta_{1,2}$ & $\theta_{3,4}$ & $\theta_{5}$ & $\theta_{6}$ & $\theta_{7}$ \\ [0.5ex] 
 \hline\hline
 (4,1) & $1.908 \pm 6.793\,i$ & $-0.091  \pm 6.793\,i$ & -5.992 & -7.992 & -9.992 \\
 \hline
 $(a,b)$ & $\theta_{1,2}$ & $\theta_{3,4}$ & $\theta_{5,6}$  & $\theta_{7}$ & $\theta_{8}$\\ [0.5ex]
 \hline\hline
 (4,2) & $1.908 \pm 6.793\,i$ &  $-0.091 \pm 6.793\,i$ & $-2.091 \pm 6.793\,i$ & -7.992 & -9.992\\
 \hline
 $(a,b)$ & $\theta_{1,2}$  & $\theta_{3,4}$ & $\theta_{5,6}$  & $\theta_{7,8}$ & $\theta_{9}$ \\ [0.5ex]
 \hline\hline
 (4,3) & $1.908 \pm 6.793\,i$ &  $-0.091 \pm 6.793\,i$ & $-2.091 \pm 6.793\,i$ & $-4.091 \pm 6.793\,i$ & -9.992 \\
\hline
  $(a,b)$ & $\theta_{1,2}$  & $\theta_{3,4}$  & $\theta_{5,6}$  & $\theta_{7,8}$  & $\theta_{9,10}$ \\ [0.5ex]
  \hline\hline
 (4,4) & $1.908 \pm 6.793\,i$ &  $-0.091 \pm 6.793\,i$ & $-2.091 \pm 6.793\,i$ & $-4.091 \pm 6.793\,i$ & $-6.091\pm 6.793\,i$\\
 \hline
 \end{tabular}
 \caption{Critical exponents in the linear-inverse parametrization.  For each truncation, a FP with two relevant directions was found in $d=4$. We refine the truncation both in $V$- and $F$-directions.}
 \label{Table3}
\end{table*}
In summary, the FPs obtained for STTs using the gauge defined by \eqref{Eq.SubSec:GFFP.3} and \eqref{Eq.SubSec:GFFP.4} feature two relevant directions for $\omega=0,1/2,1$ at least for truncations up to $(4,4)$. Since it is observed that the enlargement of the truncations in the $V$ and $F$ directions only produces more and more negative (real parts of the) new critical exponents, it is not expected that new relevant directions will be generated in the present setup. The relevant directions are associated with the gravitational couplings ${G}_k$ and ${\Lambda}_k$. At this stage, a comment is in order: Although the results presented for the $(0,0)$ truncation show good agreement with those reported in the literature, as soon as new operators are included, a discrepancy stands out. In \cite{Narain:2009fy} the corresponding FP features four relevant directions (in the linear parameterization), two of them are associated with ${G}_k$ and ${\Lambda}_k$, and the other two are related to the mass of the scalar field and the non-minimal coupling ${\xi}_2$. Such an apparent discrepancy has previously been reported in the literature, see, e.g., \cite{Eichhorn:2020sbo,Ohta:2021bkc}. However, the explicit computation of FPs and their corresponding critical exponents are subject to several different systematic uncertainties. Ideally, unphysical choices do not affect universal properties, but when approximations are made, this is not true in general, and one should look after stability of the results and favor schemes that optimize their convergence. In particular, looking at Tables \ref{Table1}, \ref{Table2} and \ref{Table3}, one sees that the real part of the critical exponents $\theta_{3,4}$ (mostly aligned with with the scalar-field mass and the non-minimal coupling $\xi_2$) are dangerously close to zero. This means that by playing with some of the unphysical choices in this computation, it is conceivable that the values of $\theta_{3,4}$ get small corrections, that are sufficient to push those critical exponents towards relevance. Of course, the opposite could also happen: apparently relevant directions can be converted into irrelevant ones by the above-mentioned systematic errors. Therefore, improving the quantitative precision is crucial to address this apparent discrepancy. In particular, even within the standard two-parameter family of standard linear covariant gauges, i.e., in the setup defined in \cite{Narain:2009fy}, fixed-point solutions featuring two or four relevant directions can be found for different choices of gauge parameters.

The gauge-fixing action \eqref{Eq.SubSec:GFFP.3} introduces an additional background-field dependence on the scalar field $\bar{\Phi}$ through the multiplicative factor $F_k (\bar{\Phi})$ and the function $f_k (\bar{\Phi})$. This dependence propagates to the ghost action, see Eq.~\eqref{Eq.SubSec:GFFP.5}. In fact, those terms break split symmetry and such spurious background dependence can render unphysical results as is well-known, see, e.g., \cite{Bridle:2013sra}. In \cite{Narain:2009fy}, the same overall factor $F_k (\bar{\Phi})$ was introduced in the gauge-fixing action but it was not included in the ghost action. In \cite{Eichhorn:2020sbo}, a systematic investigation of the effects on the inclusion of such a multiplicative factor both in the gauge-fixing as well as in the ghost actions was carried out. We have tested whether the results are sensitive to the removal of such a factor from the ghost action in the present setup. Our findings reveal that upon removal of the factor $\sqrt{F_k (\bar{\Phi})}$ from the ghost action the qualitative behavior is not affected, i.e., the FP features two relevant directions. In particular, in the linear parameterization, such a removal decreases the real part of the relevant critical exponents from ${\rm Re}~\theta_{1,2} = 1.829$ to ${\rm Re}~\theta_{1,2} = 1.611$. In the exponential parametrization, the real part of the relevant critical exponents goes from ${\rm Re}~\theta_{1,2} = 1.642$ to ${\rm Re}~\theta_{1,2} = 0.568$. In the linear-inverse parametrization, the real part of the relevant critical exponents goes from  ${\rm Re}~\theta_{1,2} = 1.908$ to ${\rm Re}~\theta_{1,2} = 0.325$. Although the qualitative properties of the fixed-point candidates are preserved in all three parameterizations, there is a substantial difference in the real part of the critical exponents in the exponential and linear-inverse choices. This is different in the linear parameterization, where these values are relatively close. Thus, one might interpret such results as a hint that the linear parameterization offers a better controlled setup since it is less sensitive to the background dependence in the ghost action.

The results reported in this section were obtained without the inclusion of the anomalous dimension of the scalar field. Such an inclusion corresponds to an improvement of the truncations considered so far and can potentially affect qualitatively the results just reported. In the next Section, we revisit the FP analysis but including the scalar-field anomalous dimension. Nevertheless, the analysis described in this section reveals evidence for the existence of a FP in STTs within a gauge that falls outside the class of gauges that was largely used in previous analysis in the literature. This provides further evidence for the existence of a suitable GMFP within STTs. Although the discrepancy in the number of relevant directions may sound alarming, it is potentially related to the lack of quantitative precision of the present truncation scheme. 

\section{Taking into account the scalar-field anomalous dimension}
\label{Sect:FP4dAD}

The FRG analysis reported in the previous section makes use of the BA. The corresponding anomalous dimensions of $h_{\mu\nu}$, $\bar{C}_\alpha$, $C^\nu$ and $\varphi$ were not included. In this section, we repeat the FP analysis by including the scalar-field anomalous dimension. As for the graviton, we leave its anomalous dimension as a free parameter and explore its impact on the fixed-point structure by setting it to different values. We neglect the anomalous dimension of the Faddeev-Popov ghosts.

In order to compute the anomalous dimensions, we set up the computation on a flat background $\bar{g}_{\mu\nu} = \delta_{\mu\nu}$, which makes it possible to apply a Fourier transform and perform the following field redefinitions,
\begin{equation}
h_{\mu\nu}\to (32\pi G_k)^{1/2} Z^{1/2}_h h_{\mu\nu}\,, \quad {\rm and} \quad \varphi \to Z^{1/2}_\varphi \varphi\,,
\label{Eq:FP4dAD.1}
\end{equation}
while for the ghosts we set $Z_c = 1$. By definition, the graviton and scalar-field anomalous dimensions are, respectively,
\begin{equation}
\eta_h = - Z^{-1}_h\partial_ t Z_h\,, \quad {\rm and} \quad \eta_\varphi = - Z^{-1}_\varphi \partial_ t Z_\varphi\,.
\label{Eq:FP4dAD.2}
\end{equation}
The anomalous dimensions are extracted from the flow of the two-point function, which is obtained from 
\begin{eqnarray}
\partial_t \Gamma^{(2)}_k &=& -\frac{1}{2}\mathrm{STr}\left(\mathbb{G}_k \Gamma^{(4)}_k \mathbb{G}_k\partial_t \mathbb{R}_k\right)\nonumber\\
&+&\mathrm{STr}\left(\mathbb{G}_k \Gamma^{(3)}_k \mathbb{G}_k \Gamma^{(3)}_k \mathbb{G}_k \partial_t \mathbb{R}_k\right)\,,
\label{Eq:FP4dAD.3}
\end{eqnarray}
Since we choose a flat background, we can use a Fourier transform to go to momentum space. In this case, we need to use the following projection to extract the scalar anomalous dimension \cite{Dona:2015tnf,Dona:2013qba,DeBrito:2019gdd},
\begin{equation}
     \eta_{\varphi}=-\frac{1}{Z_{\varphi}}\left[\frac{\partial}{\partial p^{2}}\left(\frac{\delta^{2}\partial_{t}\Gamma_{k}}{\delta\Phi(-p)\delta\Phi(p)}\right)\right]\Bigg |_{\chi=p^{2}=0}\,.
\end{equation}

Our analysis is divided in two cases: (i) We set the BA-value for the graviton anomalous dimension, i.e., $\eta_h = -2$; (ii) The graviton anomalous dimension is set to zero, $\eta_h=0$; and (iii) $\eta_h = 1.5$. Self-consistent computations of the graviton anomalous dimension indicate that the BA value is not a good approximation. Its consistency bound is $\eta_h < 2$, see \cite{Meibohm:2015twa}. Since we are treating it as a free parameter, we report on the FP structure for the BA value but consider a situation in which it is neglible and the other where it gets close to the consistency bound. Moreover, the viability of the fixed-point solution is also tested by computing the value of the scalar-field anomalous dimension at the fixed point, in order to check the consistency bound $\eta_\varphi<2$, \cite{Meibohm:2015twa}. In this section, we adopt the following truncation,
\begin{align}
    V({\Phi})&= \frac{2{\Lambda}_{k}}{16\pi{G}_{k}}+{\lambda}_{2}{\Phi}^{2}+{\lambda}_{4}{\Phi}^{4}\,,\\
     F({\Phi})&= \frac{1}{16\pi{G}_{k}}+{\xi}_{2}{\Phi}^{2}\,.
\end{align}
The expression of the scalar-field anomalous dimension at the GMFP is provided in App.~\ref{Ap:ScalarFieldAnomDim}

\subsection{$\eta_h=-2$}
\label{Sect:FP4dAD.1}

Setting the graviton anomalous dimension to its background value, i.e., $\eta_h=-2$ leads to qualitative changes for different choices for $\omega$.   In particular:
\begin{itemize}
\item Linear parameterization $(\omega = 0)$: In this case, the GMFP is given by
\begin{equation}
{\Lambda}^\ast = 0.215 \,, \quad {\rm and} \quad {G}^\ast = 0.514\,.
\end{equation}
It features four relevant directions (in contrast to the computation neglecting the anomalous dimensions). The critical exponents and the value of the scalar-field anomalous dimension at the fixed point are
\begin{eqnarray}
\theta_{1,2} &=& 2.156 \pm 3.134i\,,\quad\theta_{3,4} = 0.174 \pm 3.124i\nonumber\\
\theta_5 &=& -4.210\,,\quad \eta_{\varphi}=-0.627\,.
\end{eqnarray}
Comparing those with Table~\ref{Table1} it is clear that the inclusion of the scalar-field anomalous dimension shifts the values of $\theta_{1,2}$ and $\theta_{3,4}$ towards relevance. The shift in $\theta_{3,4}$ is enough to turn those into relevant directions. The anomalous scalar-field dimension lies within the consistency bound $\eta_\varphi < 2$. The present results compare very well with those reported in \cite{Narain:2009fy}. In particular, the critical exponents reported in \cite{Narain:2009fy} read $(\theta_{1,2};\theta_{3,4})_{\rm NP} = (2.143\pm 2.879\,i;0.143\pm 2.879\,i)$.

\item Exponential parameterization $(\omega = 1/2)$: The non-trivial values for the gravitational couplings at the GMFP are
\begin{equation}
{\Lambda}^\ast = 0.199 \,, \quad {\rm and} \quad {G}^\ast = 0.67
\,.
\end{equation}
In this case, the impact of the anomalous dimension is such that the GMFP features four relevant directions. The corresponding critical exponents and scalar-field anomalous dimension are
\begin{eqnarray}
\theta_{1,2} &=& 3.955 \pm 6.954i\,,\quad \theta_{3,4}= 2.188 \pm 6.861i\nonumber\\
\theta_5 &=& -8.650\,,\quad \eta_{\varphi}=-0.831\,.
\end{eqnarray}
Comparing those values with those reported in Table~\ref{Table2}, we identify a strong shift in the $\theta_{1,2}$ and $\theta_{3,4}$. The shift has the same qualitative effect as in the linear parameterization, i.e., driving those critical exponents to larger values. The effect in the present case is quite significant, potentially signaling a truncation artifact. The scalar-field anomalous dimension lies within the consistency bound.  

\item Linear-inverse parameterization $(\omega = 1)$: The GMFP reads
\begin{equation}
{\Lambda}^\ast = 0.125 \,, \quad {\rm and} \quad {G}^\ast =0.598\,.
\end{equation}
It features four relevant directions as in the exponential parameterization. The impact of the anomalous dimension is also very strong in this case leading to
\begin{eqnarray}
\theta_{1,2}&=&4.344 \pm 6.680i\,,\quad \theta_{3,4} = 2.700 \pm 6.486i \nonumber\\
\theta_5&=& -6.633\,,\quad \eta_{\varphi}=-1.139\,.
\end{eqnarray}
The critical exponents are rather large. This strongly suggests that this FP-candidate is not reliable.
\end{itemize}

The inclusion of the scalar-field anomalous dimension together with the background-value for the graviton anomalous dimension leads to results that can be summarized as follows:  For the linear, exponential, and linear-inverse parameterizations, the effect of such anomalous dimensions significantly enlarges the critical exponents turning $\theta_{3,4}$ into relevant directions. In addition, the absolute value of the real parts of $\theta_{3,4}$ is too large in those parameterizations if compared with the results reported in the linear parameterization and in the abscense of the anomalous dimension of the scalar-field. As already mentioned, the self-consistent graviton anomalous dimension reported in the literature differs significantly from the background value. Therefore, the present analysis is bound to be of very limited interest. However, it is worth noticing that the leading critical exponents are driven towards relevance, leading to four relevant directions. In the following, we set the graviton anomalous dimension to zero in order to get qualitatively closer to the self-consistent results.

\subsection{$\eta_h=0$}
\label{Sect:FP4dAD.2}

By setting the graviton anomalous dimension to zero, we obtain results that are qualitatively different with respect to those reported in the previous subsection. Again, we separate our analysis in three cases:

\begin{itemize}
\item Linear parameterization $(\omega = 0)$: We find the GMFP to be
\begin{equation}
{\Lambda}^\ast = 0.185 \,, \quad {\rm and} \quad {G}^\ast = 0.719\,.
\end{equation}
This fixed point features two relevant directions, differing from the previous result with $\eta_{h}=-2$. The (real parts of the) critical exponents associated with the cosmological and Newton constants remain almost the same as the ones obtained neglecting the anomalous dimension of the scalar field, see Eq.~\eqref{Eq:FP4d.2.1}. As for the critical exponents associated with the scalar mass and the non-minimal coupling, they are slightly shifted towards relevance when compared to the computation where the anomalous dimension is neglected. However, this shift is not sufficient to turn the number of relevant directions from two to four. The numerical values are
\begin{eqnarray}
\theta_{1,2} &=& 1.821 \pm 2.798i\,,\quad \theta_{3,4}= -0.156 \pm 2.787i\nonumber\\
\theta_5&=& -3.839\,,\quad\eta_{\varphi}=-0.645\,.
\end{eqnarray}
The scalar-field anomalous dimension is compatible with the consistency bound. Thus, the main qualitative features of the computation in which the scalar-field anomalous dimension is not included are preserved when it is added. This contrasts with the results for $\eta_h=-2$. 

\item Exponential parameterization $(\omega = 1/2)$: The fixed point reads
\begin{equation}
{\Lambda}^\ast = 0.190 \,, \quad {\rm and} \quad {G}^\ast = 1.02
\,.
\end{equation}
It has two relevant directions. This differs from the previous case with $\eta_h=-2$ where we have found four relevant directions in the exponential parameterization. Moreover, the same qualitative behavior observed in the case of the linear parameterization is reproduced: the critical exponents $\theta_{1,2}$ are very close to the ones obtained neglecting the anomalous dimension of the scalar field, see Eq.~\eqref{Eq:FP4d.3.1}, and $\theta_{3,4}$ are slightly shifted towards relevance but not sufficiently to turn them into relevant directions. The numerical values are
\begin{eqnarray}
\theta_{1,2}&=& 1.500 \pm 6.424i\,,\quad \theta_{3,4}= -0.275 \pm 6.427i\nonumber\\
\theta_5 &=& -7.542\,,\quad \eta_{\varphi}=-0.594\,.
\end{eqnarray}
The scalar-field anomalous dimension remains within the consistency bound, a property that is slightly different with respect to the results reported in \cite{Ohta:2021bkc}, where a fixed point with two relevant directions is found, but the scalar-anomalous dimension is rather large in the exponential parameterization, casting doubts on the robustness of such a solution as the authors point out.

\item Linear-inverse parameterization $(\omega = 1)$: In this case, we found the following fixed-point solution,
\begin{equation}
{\Lambda}^\ast = 0.118 \,, \quad {\rm and} \quad {G}^\ast = 0.975\,.
\end{equation}
It features four relevant directions, a fact that agrees with the result reported for $\eta_h=-2$. However, unlike the previous case, the critical exponents are not strongly modified with respect to those associated with the computation where the anomalous dimensions are neglected. Moreover, in the case of $\eta_h=-2$, the scalar-field anomalous dimension was outside the consistency interval, suggesting that such a fixed point would be a truncation artifact. For $\eta_h=0$, the anomalous dimension of the scalar field albeit large in absolute value is compatible with the consistency bound. The numerical values are
\begin{eqnarray}
\theta_{1,2}&=& 1.642 \pm 6.868i\,,\quad \theta_{3,4}= 0.072 \pm 6.833i\nonumber\\
\theta_5&=&-6.041\,,\quad\eta_{\varphi}= -0.861\,.
\end{eqnarray}
The qualitative features of this fixed point are in agreement with works such as \cite{Narain:2009fy,Eichhorn:2020sbo} where four relevant directions are found. However, to our knowledge, this is the first time that an explicit computation is carried out employing a linear split of the inverse metric (and, additionally, within the mFDS gauge).
\end{itemize}

The fixed-point structure for different parameterizations neglecting the graviton anomalous dimension but taking into account the one associated with the scalar field can be summarized as follows: The results are in qualitative agreement with those obtained within the BA for linear and exponential parameterizations. As for the linear-inverse parameterization, the fixed-point features two additional relevant directions, i.e., it has a total of four relevant directions associated with the Newton and cosmological constants, the scalar mass and the non-minimal coupling. In all cases, the absolute value of the real part of $\theta_{3,4}$ is very close to zero and establishing whether it is positive or negative requires quantitative precision. However, although discrepant signs can be found within different schemes, the corresponding absolute values are not very far apart, which can be interpreted as a signal that the converged value is very small indeed. This concretely shows that, in the present case, due to the smallness of the real part of $\theta_{3,4}$, unphysical choices such as field parameterization can lead to an apparent discrepancy in the number of relevant directions. In the following, we take the anomalous dimension of the graviton closer to the consistency bound.

\subsection{$\eta_h=1.5$}
\label{Sect:FP4dAD.3}

\begin{itemize}
\item Linear parameterization $(\omega = 0)$: The GMFP is characterized by
\begin{equation}
{\Lambda}^\ast =  0.133 \,, \quad {\rm and} \quad {G}^\ast = 1.024\,.
\end{equation}
It features two relevant directions at in the case where $\eta_h=0$. The value of the Newton constant at the fixed point has a significant increase with respect to the $\eta_h=0$ result. The critical exponents and the value of the anomalous dimension of the scalar field at the fixed point are close to the numbers reported for $\eta_h=0$, 
\begin{eqnarray}
\theta_{1,2} &=&  1.594\pm 2.209i\,,\quad \theta_{3,4}= -0.381 \pm 2.198i\nonumber\\
\theta_5&=& -3.289\,,\quad\eta_{\varphi}=-0.649\,.
\end{eqnarray}
Hence, in the linear parameterization, the FP structure in the case where the anomalous dimension of the graviton is taken closer to the consistency bound is essentially the same as those with $\eta_h=0$ qualitatively and even quantitatively.

\item Exponential parameterization $(\omega = 1/2)$: The fixed point reads
\begin{equation}
{\Lambda}^\ast = 0.163 \,, \quad {\rm and} \quad {G}^\ast = 1.789
\,.
\end{equation}
There is a significant increase in the value of Newton's coupling fixed point with respect to the result for $\eta_h=0$. The FP features two relevant directions but the critical exponents are rather different with respect to the $\eta_h = 0$ case. In particular, the critical exponents associated with the gravitational couplings are severely lowered while the critical exponents $\theta_{3,4}$ have their real parts substantially increased in absolute value. This indicates that treating the anomalous dimension as a free parameter in this case is not an efficient and trustable strategy to grasp the essential features of the FP structure. The critical exponents read
\begin{eqnarray}
\theta_{1,2}&=& 0.056 \pm 4.063i\,,\quad \theta_{3,4}= -1.800 \pm 4.141i\nonumber\\
\theta_5 &=& -5.179\,,\quad \eta_{\varphi}=-0.257\,.
\end{eqnarray}
Notice that the anomalous dimension of the scalar field remains within the consistency bound but it is substantially smaller in absolute value with respect to the previous analysis in the exponential parameterization. 

\item Linear-inverse parameterization $(\omega = 1)$: In this case, the FP-candidate is
\begin{equation}
{\Lambda}^\ast = 0.101 \,, \quad {\rm and} \quad {G}^\ast = 1.964\,.
\end{equation}
The situation is similar to the one reported in the exponential parameterization but with a key difference: The shift in the critical exponents is so severe that those associated with the gravitational couplings become irrelevant, 
\begin{eqnarray}
\theta_{1,2}&=& -1.794 \pm 4.965i\,,\quad \theta_{3,4}= -3.287 \pm 5.337i\nonumber\\
\theta_5&=&-4.555\,,\quad\eta_{\varphi}= -0.332\,.
\end{eqnarray}
This shows that pushing the anomalous dimension closer to the consistency bound leads to a drastic change with respect to $\eta_h = 0$. In particular, by progressively increasing $\eta_h$, we observe this flip in the sign of the real part of $\theta_{1,2}$ at $\eta_h \approx 0.8$. 
\end{itemize}

This section reveals that, for values of the anomalous dimension of the graviton within the consistency bound but far from the BA value, the FP candidates tend to have two relevant directions. The linear parameterization shows reasonable stability with respect to different choices of $\eta_h$. As for the exponential and linear-inverse parameterizations, we have detected a severe dependence of the critical exponents on the choice of $\eta_h$. At this stage, a comment is in order: The treatment we gave to the graviton anomalous dimension is not self-consistent. In particular, $\eta_h$ depends on the other couplings and, when computing the stability matrix, the derivative of the beta functions should hit the expression of $\eta_h$. This is not accounted for in the present setup since $\eta_h$ is taken as a fixed parameter from the outset. It is not clear if a self-consistent treatment would lead to qualitatively different results and this deserves further investigation. From what we have reported, a tentative interpretation is that such an impact should be more modest in the linear parameterization where we observe reasonably stable results. In the next Section, we perform a parametric analysis in order to chart the region in the gravitational couplings space that render $\theta_{3,4}$ (ir)relevant directions. Notably, the critical exponent $\theta_5$, associated with the quartic coupling $\lambda_4$, always corresponds to an irrelevant direction.

\section{Parametric Analysis} \label{Sect:ParametricAnalysis}

The results reported in the previous sections indicate that suitable fixed-point candidates might feature two or four relevant directions within the approximations employed in this work. In this section we analyze the critical exponents $\theta_3$ and $\theta_4$ in more details. In particular, instead of looking at the sign of (the real parts of) those critical exponents at the exact FP value obtained within the truncations employed in this work, we assume that a GMFP exists and treat the fixed-point value for $G$ and $\Lambda$ as free parameters. Additionally, we attribute different values for the graviton anomalous dimension. For each choice of parameterization, i.e., for linear $(\omega=0)$, exponential $(\omega=1/2)$ and linear-inverse $(\omega=1)$ metric parameterizations, we compute the critical exponents $\theta_{3,4}$. The results are reported in several plots below.

An operationally useful expression for the critical exponents $\theta_{3,4}$ can be obtained as follows: We consider the five-coupling truncation involving $\mathbf{g} = (G,\Lambda,\lambda_2,\xi_2,\lambda_4)$ and a GMFP is characterized by $\mathbf{g}^{\ast} = (G^\ast,\Lambda^\ast,0,0,0)$. The stability matrix $\mathbb{M}$ has the following structure,
\begin{eqnarray}
\mathbb{M} &=& \left(
\begin{matrix}
\partial_{G} \beta_G & \partial_{\Lambda} \beta_G & \partial_{\lambda_2} \beta_{G} & \partial_{\xi_2} \beta_{G} & \partial_{\lambda_4}  \beta_{G}\\
\partial_{G} \beta_\Lambda & \partial_{\Lambda} \beta_\Lambda & \partial_{\lambda_2} \beta_{\Lambda} & \partial_{\xi_2} \beta_{\Lambda} & \partial_{\lambda_4}  \beta_{\Lambda}\\
0 & 0 & \partial_{\lambda_2} \beta_{\lambda_2} & \partial_{\xi_2} \beta_{\lambda_2} & \partial_{\lambda_4}  \beta_{\lambda_2}\\
0 & 0 & \partial_{\lambda_2} \beta_{\xi_2} & \partial_{\xi_2} \beta_{\xi_2} & \partial_{\lambda_4}  \beta_{\xi_2}\\
0 & 0 & \partial_{\lambda_2} \beta_{\lambda_4} & \partial_{\xi_2} \beta_{\lambda_4} & \partial_{\lambda_4}  \beta_{\lambda_4}\\
\end{matrix}
\right)_{\mathbf{g}=\mathbf{g}^\ast}\nonumber\\
&=& \left(
\begin{matrix}
\mathcal{A}_{\rm grav} & \mathcal{C} \\
0 & \mathcal{B}_{\rm matter}
\end{matrix}
\right)\,.
\label{Eq:StabTrian}
\end{eqnarray}
The lower-left block vanishes at the GMFP. This can be understood as follows: Consider the beta functions of the matter couplings $\beta_a$ with $a=(\lambda_2,\xi_2,\lambda_4)$. Around the Gaussian fixed point in the matter sector, one has
\begin{eqnarray}
\beta_a (G,\Lambda;\lambda_2,\xi_2,\lambda_4) &\approx& \beta_a (G,\Lambda; 0,0,0) + A_a (G,\Lambda) \lambda_2 \nonumber\\
&+& B_a (G,\Lambda) \xi_2 + C_a (G,\Lambda) \lambda_4\,,\nonumber\\
\label{Eq:GenBetaStruc}
\end{eqnarray}
with
\begin{equation}
\left(A_a,B_a,C_a\right)\equiv\left( \frac{\partial \beta_a}{\partial \lambda_2}
,\frac{\partial \beta_a}{\partial \xi_2},\frac{\partial \beta_a}{\partial \lambda_4}\right)_{a^\ast = 0}\,.
\end{equation}
Hence, the derivative of $\beta_a$ with respect to $G$ or $\Lambda$ always has linear terms in the matter couplings. Thus, at the GMFP, those couplings vanish so as the derivatives of $\beta_a$ with respect to $G$ and $\Lambda$. This leads to the vanishing of the lower-left block of $\mathbb{M}$. The upper-right block denoted by $\mathcal{C}$ does not vanish in general. However, due to the block-triangular structure of \eqref{Eq:StabTrian}, the matrix $\mathcal{C}$ does not affect the eigenvalues of $\mathbb{M}$. Therefore, in order to explore the critical exponents $\theta_{3,4}$ we can restrict our attention to $\mathcal{B}_{\rm matter}$. At this stage, a comment is in order: Near the Gaussian fixed point in the matter sector, the beta function associated with $\lambda_4$ has the following structure,
\begin{eqnarray}
\beta_{\lambda_4} (G,\Lambda;\lambda_2,\xi_2,\lambda_4) &=& \beta_{\lambda_4} (G,\Lambda;0,0,0) + C_{4} (G,\Lambda) \lambda_4 \nonumber\\
&+& \mathcal{O}(g^2)\,,
\end{eqnarray}
with 
\begin{equation}
C_{4} (G,\Lambda) = \frac{\partial \beta_{\lambda_4}}{\partial \lambda_4}\Bigg|_{a^\ast = 0}\,.
\end{equation}
The coefficients $A_{4} (G,\Lambda)$ and $B_4 (G,\Lambda)$, see Eq.~\eqref{Eq:GenBetaStruc}, vanish because they would correspond to linear insertions of couplings that are associated with field operators that are quadratic in the scalar field. A quartic-order interaction is not generated by the insertion of a single two-fields interaction vertex. Those couplings enter, at least, at quadratic order in $\beta_{\lambda_4}$ and thus its derivative with respect to $(\lambda_2,\xi_2)$ at the Gaussian fixed point vanishes. Hence, at the Gaussian fixed point,
\begin{equation}
\partial_{\lambda_2}\beta_{\lambda_4} = \partial_{\xi_2} \beta_{\lambda_4} = 0\,.
\end{equation}
Therefore, the matrix $\mathcal{B}_{\rm matter}$ has the following structure,
\begin{equation}
\mathcal{B}_{\rm matter} = \left(
\begin{matrix}
\partial_{\lambda_2} \beta_{\lambda_2} & \partial_{\xi_2} \beta_{\lambda_2} & \partial_{\lambda_4}  \beta_{\lambda_2}\\
\partial_{\lambda_2} \beta_{\xi_2} & \partial_{\xi_2} \beta_{\xi_2} & \partial_{\lambda_4}  \beta_{\xi_2}\\
0 & 0 & \partial_{\lambda_4}  \beta_{\lambda_4}\\
\end{matrix}
\right)_{\mathbf{g}=\mathbf{g}^\ast}\,.
\end{equation}
Once again, we end up with a block-tringular matrix. Thus, the critical exponents associated with the matter sector are extracted by
\begin{equation}
\theta_{3,4} = - {\rm Eig}\left(
\begin{matrix}
\partial_{\lambda_2} \beta_{\lambda_2} & \partial_{\xi_2} \beta_{\lambda_2} \\
\partial_{\lambda_2} \beta_{\xi_2} & \partial_{\xi_2} \beta_{\xi_2}
\end{matrix}
\right)_{\mathbf{g}=\mathbf{g}^\ast}\equiv -{\rm Eig}~\mathcal{B}^{(2)}_{\rm matter}\,, 
\label{Eq:CritExpMatterBlockMatrices}
\end{equation}
and
\begin{equation}
\theta_5 = - \frac{\partial  \beta_{\lambda_4}}{\partial \lambda_4}\Bigg|_{\mathbf{g} = \mathbf{g}^\ast}\,.
\label{Eq:CritExpMatterBlockMatrices.2}
\end{equation}
Our concern from now on is to evaluate $\theta_{3,4}$ from Eq.~\eqref{Eq:CritExpMatterBlockMatrices}. The critical exponents $\theta_{3,4}$ are given by
\begin{equation}
\theta_{3,4} = -\frac{1}{2}{\rm Tr}~\mathcal{B}^{(2)}_{\rm matter} \mp \frac{1}{2}\sqrt{\Delta}\,,
\end{equation}
with
\begin{equation}
\Delta = \left({\rm Tr}~\mathcal{B}^{(2)}_{\rm matter}\right)^2 - 4\,{\rm det}~\mathcal{B}^{(2)}_{\rm matter}\,.
\end{equation}
For $\Delta > 0$, the critical exponents $\theta_{3,4}$ are distinct and real; for $\Delta = 0$, they are equal and real and for $\Delta < 0$, they form a complex conjugate pair.

\begin{figure*}[t!]
    \centering
    \includegraphics[width=1.0\linewidth]{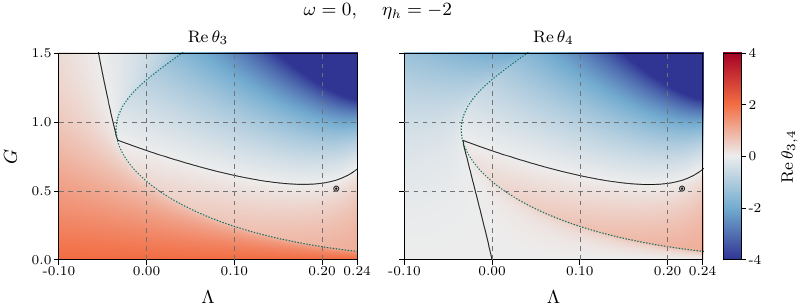}
    \includegraphics[width=1.0\linewidth]{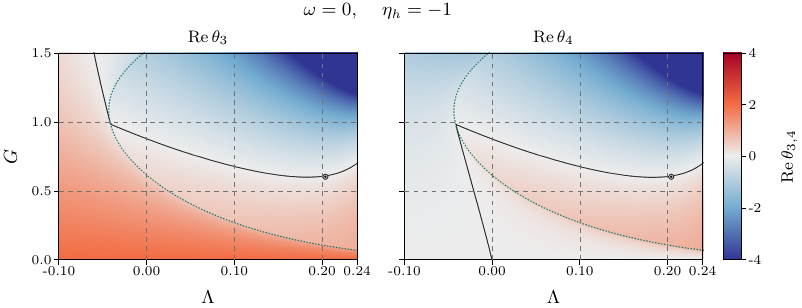}
    \includegraphics[width=1.0\linewidth]{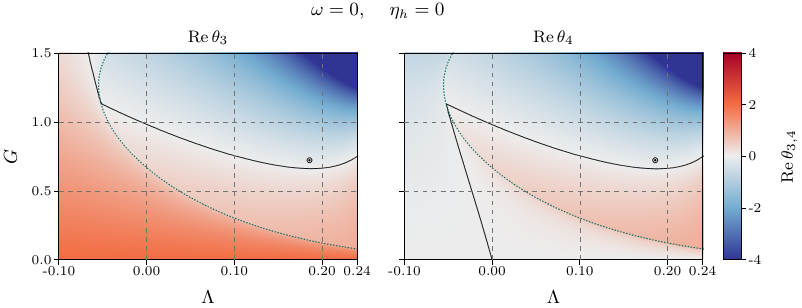}
    \caption{Real parts of $\theta_3$ and $\theta_4$ in the linear parameterization $\omega=0$. The plots are displayed for different choices of the graviton anomalous dimension $\eta_h$. The solid black line corresponds to ${\rm Re}~\theta_{3,4}=0$ and the dashed green line divides the region where the critical exponents are complex and real. The bound $\eta_\varphi < 2$ is satisfied throughout the displayed region. The small black-white circle represents the actual value of the fixed-point for the corresponding choice of parameters.}
    \label{Fig:LinParTheta34_1}
\end{figure*}
\begin{figure*}[t!]
    \centering
    \includegraphics[width=1.0\linewidth]{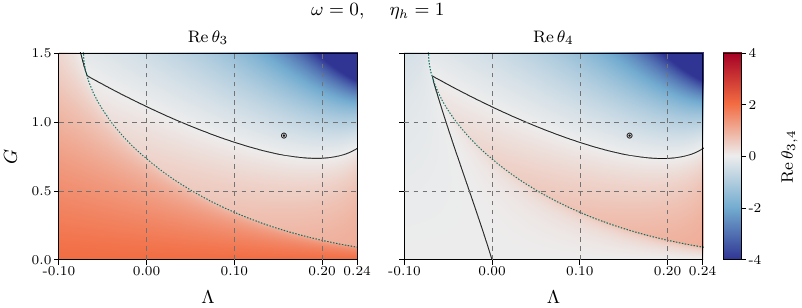}
    \includegraphics[width=1.0\linewidth]{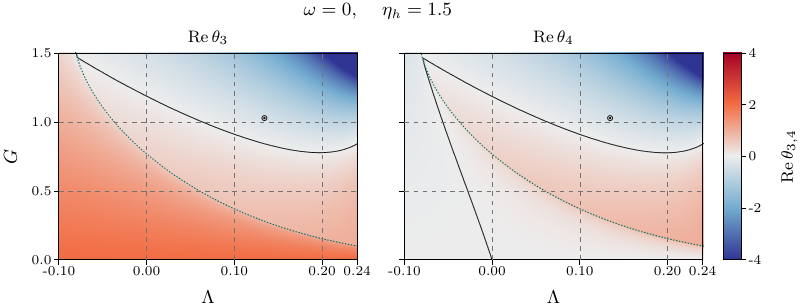}
    \caption{Real parts of $\theta_3$ and $\theta_4$ in the linear parameterization $\omega=0$. The plots are displayed for different choices of the graviton anomalous dimension $\eta_h$. The solid black line corresponds to ${\rm Re}~\theta_{3,4}=0$ and the dashed green line divides the region where the critical exponents are complex and real. The bound $\eta_\varphi < 2$ is satisfied throughout the displayed region. The small black-white circle represents the actual value of the fixed-point for the corresponding choice of parameters.}
    \label{Fig:LinParTheta34_2}
\end{figure*}

Now, we have a concrete expression for the critical exponents $\theta_{3,4}$ that involves the interpolating parameter $\omega$, the gravitational couplings $G$ and $\Lambda$ as well as the graviton anomalous dimension $\eta_h$. In the previous sections, we have solved fixed-point equations and found the values $(G^\ast,\Lambda^\ast)$ for different choices of the interpolating parameter and the graviton anomalous dimension. In the present section, we treat the gravitational couplings as free parameters and collect plots of the critical exponents $\theta_{3,4}$ for different choices of $\eta_h$. By treating the gravitational couplings as free parameters, we are assuming that a GMFP exists. In Fig.~\ref{Fig:LinParTheta34_1} and \ref{Fig:LinParTheta34_2} we show the plots of the real parts of $\theta_{3,4}$ within the linear parameterization. Those associated with the exponential and linear-inverse parameterizations are omitted due to the strong dependence on $\theta_{1,2}$ on the value of the anomalous dimension of the graviton.

There is a clear trend in all three parameterizations (which is can be understood from the plots in the linear parameterization): Starting from the BA graviton-anomalous dimension $(\eta_h=-2)$ and increasing it up $\eta_h=1.5$, the fixed point crosses the ${\rm Re}~\theta_{3,4}=0$ line moving from a region with four relevant directions to two relevant directions. The crossing occurs between $\eta_h=-1$ and $\eta_h=0$ for $\omega=0$ and $\omega=1/2$ while, for $\omega=1$, the crossing occurs for $\eta_h$ slightly larger than zero. The fixed point is always sufficiently away from the regulator bound $\eta_\varphi<2$. Interestingly, for sufficiently small (or negative) values of the cosmological constant fixed-point, a wide region with real $\theta_{3,4}$ critical exponents opens up, leaving room for fixed points with three relevant directions. In all cases, the fixed point is dangerously close to the ${\rm Re}~\theta_{3,4}=0$ line. This reflects the difficulty in characterizing the number of relevant directions: Any minor change arising from unphysical choices such as regulator shape function, gauge and field parameterization can move the fixed point across the ${\rm Re}~\theta_{3,4}=0$. This analysis corroborates what we have seen before: The real parts of $\theta_{3,4}$ are too close to zero indicating that quantitative precision is indispensable to establish the number of relevant directions in STTs.

\section{Conclusion}

The use of FRG techniques to search for suitable fixed points in quantum gravity requires a comprehensive exploration of the underlying systematic uncertainties behind the underlying approximations. The refinement of truncations should reveal (apparent) convergence of universal properties of the would-be fixed-point if it really exists. Yet beta functions are not universal themselves and carry spurious dependences from, e.g., gauge choices and field parameterization. Hence, different computational schemes can lead to significantly different beta functions, although the universal properties of a true fixed point should be preserved (within systematic uncertainties) against such choices. Since explicit computations involve truncations, stability of results can get fogged by truncation artifacts. 

One convenient strategy to test such stability-properties is to leave free parameters in the corresponding unphysical ingredients of the computation, for instance, gauge condition, field parameterization and choice of regulator. The resulting beta functions will depend on such parameters and, by setting different values to those, an explicit check for stability can be performed. Clearly, if a fixed point exists, it is very well conceivable that there is a particular choice for such parameters that speed up the convergence of the results for truncation refinements or, in other words, that optimize the computations. 

In this work, we have explored the impact of choosing a gauge condition that is not connected to the standard two-parameter family of linear covariant gauges mostly employed in the ASQG literature. It involves the dynamical scalar field itself and is constructed in such a way to eliminate non-minimal terms in the Hessian. As such, we can carry out the computation without the use of the so-called York decomposition allowing for an investigation of the fixed-point structure within a different class of gauge and different choice of fields that are used in the coarse-graining procedure. We also left a free parameter in the field parameterization of metric fluctuations. 

It is expected that, if a fixed point does not exist, then by changing all those elements in the explicit computation of the beta functions in a given truncation, one can get very different results with respect to the ones already known in the literature. Yet we have verified that, although some qualitative differences appear, the fixed-point structure that is known in the literature is still found. Additionally, by adjusting the different arbitrary choices suitably, we re-obtain qualitative features that are in agreement with results reported in the literature. This agrees with the intuition that changes on unphysical elements in a given truncation can more or less compensate for each other to preserve the would-be universal properties if any. 

Our findings show that, within the truncations employed, a fixed-point candidate with two or four relevant directions emerge. As pointed out, the apparent discrepancy in the number of free parameters lies on the very small (real part of the) critical exponents associated with the scalar mass and non-minimal coupling. It is necessary to improve the present computation to establish whether those directions remain irrelevant or relevant. However, such apparent contradicting result is not new and reported in several works in the literature. Perhaps, what is most important is the fact that for every choice of field parameterization, we have found a suitable fixed-point candidate in the mFDS gauge corroborating the findings reported in the literature within standard linear covariant gauges together with the York decomposition. This corresponds to a non-trivial check for the robustness of the fixed point in STTs. One particular weakness of our analysis is the way we include the anomalous dimension of the graviton. Nevertheless, we observed a reasonable stable behavior of the fixed-point structure in the linear parameterization which might indicate that this choice speeds up the convergence of the properties of the would-be fixed point.Once we treated the graviton anomalous dimension as a free parameter, we could analyze how it affects the signal of the second pair of critical exponents. Within this scenario, we found that the larger the graviton anomalous dimension, the more negative the real part of $\text{Re};\theta_{3,4}$ becomes. This behavior led to a change in the number of relevant directions from four to two. As a future possible analysis, we could see how the inclusion of matter fields and their respective anomalous dimensions may affect the results. Also, the explicit computation of the graviton anomalous dimension may shed light on the numbers of relevant directions. 

\section*{Acknowledgments}

We are grateful to Gustavo P. de Brito and Astrid Eichhorn for enlightening discussions. ADP acknowledges support from CNPq under the grant PQ-B (307623/2026-2) and from FAPERJ through the “Jovem Cientista do Nosso Estado” program (E-26/204.457/2025).

\onecolumngrid 
\appendix

    \section{Conventions and Expansion of Curvature Tensors}
\label{Ap:GeometricO}

The analysis of this work is carried out for $d$-dimensional Euclidean metrics. The metric is expanded as $g_{\mu\nu}=\bar{g}_{\mu\nu}+\delta g_{\mu\nu}$ with
\begin{equation}
\delta g_{\mu\nu} = h_{\mu\nu} + \omega {h_\mu}^{\rho}{h^{\rho}}_{\nu}\,,
\label{setup4}
\end{equation}
where $\bar{g}_{\mu\nu}$ is a fixed background metric and $h_{\mu\nu}$ corresponds to the fluctuations that are not assumed to be small. For the scalar field, we split it as $\Phi = \bar{\Phi}+\varphi$ with a fixed background $\bar{\Phi}$. Another useful formulae that are used throughout the work are,
\begin{equation}
g^{\mu\nu} = \bar{g}^{\mu\nu}-h^{\mu\nu}+(1-\omega){h^{\mu}}_{\rho}{h^{\rho\nu}}+\mathcal{O}(h^3)\,,
\label{setup5.1}
\end{equation}
\begin{equation}
\sqrt{g} = \sqrt{\bar{g}}+\sqrt{\bar{g}}\frac{1}{2}h + \sqrt{\bar{g}}\frac{1}{2}\left(\omega-\frac{1}{2}\right){h^{\mu}}_{\rho}{h^{\rho}}_{\mu}+\sqrt{\bar{g}}\frac{1}{8}h^2+\mathcal{O}(h^3)\,,
\label{setup6}
\end{equation}
\begin{equation}
R_{\beta\nu} = \bar{R}_{\beta\nu}+R^{(1)}_{\beta\nu}+R^{(2)}_{\beta\nu}+\mathcal{O}(h^3)\,,
\label{setup7}
\end{equation}
with
\begin{eqnarray}
R^{(1)}_{\beta\nu} = -\frac{1}{2}\left(\bar{\nabla}^2 h_{\beta\nu}+\bar{\nabla}_{\beta}\bar{\nabla}_\nu h-\bar{\nabla}_{\beta}\bar{\nabla}_\alpha {h^{\alpha}}_\nu-\bar{\nabla}_{\nu}\bar{\nabla}_\alpha {h^{\alpha}}_\beta\right)+\frac{1}{2}\bar{R}_{\beta\alpha}{h^{\alpha}}_{\nu}+\frac{1}{2}\bar{R}_{\nu\alpha}{h^{\alpha}}_{\beta}
-\bar{R}_{\beta\alpha\nu\lambda}h^{\alpha\lambda}\,,
\label{setup8}
\end{eqnarray}
\begin{eqnarray}
R^{(2)}_{\beta\nu} &=& \frac{\omega}{2}\bar{\nabla}_\mu\Big[\bar{\nabla}_\beta\left(h_{\nu\sigma}h^{\sigma\mu}\right)+\bar{\nabla}_\nu\left(h_{\beta\sigma}h^{\sigma\mu}\right)-\bar{\nabla}^\mu\left(h_{\beta\sigma}{h^{\sigma}}_\nu\right)\Big]-\left(\omega-\frac{1}{2}\right)\bar{\nabla}_\nu\left(h_{\mu\sigma}\bar{\nabla}_\beta h^{\sigma\mu}\right)\nonumber\\
&-&\frac{1}{2}\bar{\nabla}_\mu\Big[h^{\mu\lambda}\left(\bar{\nabla}_\beta h_{\nu\lambda}+\bar{\nabla}_\nu h_{\beta\lambda}-\bar{\nabla}_\lambda h_{\beta\nu}\right)\Big]+\frac{1}{4}\bar{\nabla}_\sigma h\left(\bar{\nabla}_\beta {h^{\sigma}}_{\nu}+\bar{\nabla}_\nu {h^{\sigma}}_{\beta}-\bar{\nabla}^\sigma h_{\beta\nu}\right)\nonumber\\
&-&\frac{1}{4}\left(\bar{\nabla}_\sigma {h^{\mu}}_\nu+\bar{\nabla}_\nu {h^{\mu}}_\sigma-\bar{\nabla}^{\mu} h_{\nu\sigma}\right)\left(\bar{\nabla}_\beta{h^{\sigma}}_{\mu}+\bar{\nabla}_\mu{h^{\sigma}}_{\beta}-\bar{\nabla}^\sigma h_{\mu\beta}\right)\,.
\label{setup9}
\end{eqnarray}
For the Ricci scalar,
\begin{equation}
R = \bar{R}+R^{(1)}+R^{(2)}+\mathcal{O}(h^3)\,,
\label{setup10}
\end{equation}
with
\begin{equation}
R^{(1)} = \bar{\nabla}_\mu\bar{\nabla}_\nu h^{\mu\nu} - \bar{\nabla}^2 h - h^{\mu\nu}\bar{R}_{\mu\nu}\,,
\label{setup11}
\end{equation}
\begin{eqnarray}
R^{(2)} &=& \bar{\nabla}_\beta h^{\beta\lambda}\bar{\nabla}_\lambda h + h^{\beta\lambda}\bar{\nabla}_\beta \bar{\nabla}_\lambda h - \frac{1}{4}\bar{\nabla}_\lambda h \bar{\nabla}^\lambda h-\left(2\omega-\frac{3}{4}\right)\bar{\nabla}^\beta {h^{\mu}}_\lambda \bar{\nabla}_{\beta}{h^{\lambda}}_{\mu}
- (2\omega - 1)h_{\beta\lambda}\bar{\nabla}^2 h^{\lambda\beta}\nonumber\\
&+&(\omega-1)\bar{\nabla}_\beta h^{\lambda\beta}\bar{\nabla}^\mu h_{\mu\lambda}
+2(\omega-1)h^{\lambda\beta}\bar{\nabla}_\beta\bar{\nabla}^{\mu}h_{\mu\lambda}
+\left(\omega-\frac{1}{2}\right)\bar{\nabla}_\beta h_{\mu\sigma}\bar{\nabla}^\mu h^{\sigma\beta}+(\omega-1)\bar{R}_{\sigma\lambda\beta\mu}h^{\lambda\beta}h^{\mu\sigma}\,.
\label{setup12}
\end{eqnarray}

\section{Second Order Contributions in the Fluctuations Fields}
\label{Ap:Fluctuations}

The terms involving the potential $V_k (\bar{\Phi})$ are given by

\begin{equation}
\Gamma^{(V)}_k = \int {\rm d}^dx \sqrt{\bar{g}}\left[\frac{1}{2}V^{\prime \prime}_k (\bar{\Phi})\varphi^2 + \frac{1}{2} h V^\prime_k (\bar{\Phi})\varphi + \frac{1}{2}\left(\omega-\frac{1}{2}\right){h^\mu}_{\nu}{h^\nu}_{\mu} V_k (\bar{\Phi})+\frac{1}{8}h^2 V_k (\bar{\Phi})\right]
\label{fluctuationV.1}
\end{equation}

The terms involving the potential $G_k (\bar{\Phi})$ (where we take the background field $\bar{\Phi}$ to be constant,

\begin{equation}
\Gamma^{(G)}_k = \int {\rm d}^dx\sqrt{\bar{g}}\left[\frac{1}{2}\bar{g}^{\mu\nu}\partial_\mu \varphi \partial_\nu \varphi\right]
\label{fluctuationG.1}
\end{equation}

The terms involving the potential $F_k (\bar{\Phi})$

\begin{eqnarray}
\Gamma^{(F)}_k &=&\int {\rm d}^dx\sqrt{\bar{g}}F_k (\bar{\Phi})\left[-\bar{\nabla}_\beta h^{\beta\lambda}\bar{\nabla}_\lambda h - h^{\beta\lambda}\bar{\nabla}_{\beta}\bar{\nabla}_\lambda h + \frac{1}{4}\bar{\nabla}_\lambda h \bar{\nabla}^\lambda h + \left(2\omega - \frac{3}{4}\right)\bar{\nabla}^{\beta}{h^{\mu}}_{\lambda}\bar{\nabla}_\beta {h^{\lambda}}_{\mu}+(2\omega-1)h_{\beta\lambda}\bar{\nabla}^2 h^{\lambda\beta}\right.\nonumber\\
&-&\left.(\omega-1)\bar{\nabla}_\beta h^{\lambda\beta}\bar{\nabla}^{\mu}h_{\mu\lambda}-2(\omega-1)h^{\lambda\beta}\bar{\nabla}_\beta \bar{\nabla}^\mu h_{\mu\lambda}-\left(\omega-\frac{1}{2}\right)\bar{\nabla}_\beta h_{\mu\sigma}\bar{\nabla}^{\mu} h^{\sigma\beta}-\frac{1}{2}h\bar{\nabla}_\mu\bar{\nabla}_\nu h^{\mu\nu}+\frac{1}{2}h\bar{\nabla}^2 h\right.\nonumber\\
&-&\left.(\omega-1)\bar{R}_{\sigma\lambda\beta\mu}h^{\lambda\beta}h^{\mu\sigma}
+\frac{1}{2}hh_{\mu\nu}\bar{R}^{\mu\nu}-\left(\frac{1}{2}\left(\omega - \frac{1}{2}\right){h^\mu}_\nu {h^{\nu}}_{\mu}+\frac{1}{8}h^2\right)\bar{R}\right]\nonumber\\
&-&\int {\rm d}^dx \sqrt{\bar{g}}F^\prime_k (\bar{\Phi})\left[\varphi\bar{\nabla}_\mu \bar{\nabla}_\nu h^{\mu\nu} -\varphi \bar{\nabla}^2 h - \varphi h^{\mu\nu}\bar{R}_{\mu\nu}+\frac{1}{2}\varphi h \bar{R}\right]\nonumber\\
&-& \int {\rm d}^dx \sqrt{\bar{g}}F^{\prime\prime}_k (\bar{\Phi})\frac{1}{2}\varphi^2 \bar{R}.
\label{fluctuationF.1}
\end{eqnarray}

\section{Scalar-field Anomalous Dimension}
\label{Ap:ScalarFieldAnomDim}

At the GMFP, the anomalous dimension of the scalar field is given by
\begin{equation}
    \eta_\varphi(G,\Lambda,\eta_h,\omega)=
\frac{G}{12\pi}\frac{
\left[
\dfrac{1}{\gamma(\omega)}-\dfrac{9}{\tilde{\gamma}(\omega)}
+\left(1-6\omega+\dfrac{8\omega-1}{8}\eta_h\right)
\left(\dfrac{1}{\gamma^2(\omega)}-\dfrac{9}{\tilde{\gamma}^2(\omega)}\right)
\right]}
{1+\dfrac{G}{96\pi}\left[\dfrac{1}{\gamma (\omega)}-\dfrac{9}{\tilde{\gamma}(\omega)}\right]}\,,
\end{equation}
with $\gamma (\omega)\equiv 1-2(1+2\omega)\Lambda$ and
$\tilde{\gamma}(\omega)\equiv 1+2(2\omega-1)\Lambda$. In particular,
\begin{itemize}
    \item $\omega = 0$

\begin{equation}
\eta_\varphi\Big|_{\omega=0}=
\frac{G\left(-16+\eta_h+16\Lambda\right)}
{(-1+2\Lambda)\left(G-12\pi+24\pi\Lambda\right)}\,.
\label{eq:eta-linear}
\end{equation}

    \item $\omega = 1/2$

\begin{equation}
\eta_\varphi\Big|_{\omega=1/2}=
\frac{2G\left(
-8+3\eta_h+76\Lambda-27\eta_h\Lambda
-144\Lambda^2+54\eta_h\Lambda^2
\right)}
{(-1+4\Lambda)
\left(-2G+9G\Lambda+24\pi-96\pi\Lambda\right)}\,.
\label{eq:eta-exponential}
\end{equation}

    \item $\omega=1$

\begin{eqnarray}
\eta_\varphi\Big|_{\omega=1}
=-\frac{G\left(
-32+7\eta_h+472\Lambda-98\eta_h\Lambda
-1472\Lambda^2+280\eta_h\Lambda^2+672\Lambda^3
\right)}
{(1+2\Lambda)(-1+6\Lambda)(G-7G\Lambda-12\pi+48\pi\Lambda
+144\pi\Lambda^2)}\,.
\label{eq:eta-inverse}
\end{eqnarray}

\end{itemize}
\clearpage

\clearpage

\twocolumngrid
\bibliography{refs}

\end{document}